\documentclass[twocolumn,showpacs,preprintnumbers,amsmath,amssymb,prc,nofootinbib,final]{revtex4-1}

\usepackage{graphicx}
\usepackage{dcolumn}
\usepackage{bm}
\usepackage{multirow}
\usepackage{color}
\usepackage{amsmath}
\usepackage{mathtools}
\usepackage{textcomp}
\usepackage{amssymb}
\usepackage{hyperref}

\usepackage[caption=false]{subfig}
\usepackage{dcolumn}
\usepackage{natbib}
\usepackage{color}
\usepackage[normalem]{ulem}
\usepackage{soul,xcolor}
\usepackage{xfrac}
\usepackage{cancel}

\newcommand{\nuc}[2]{$^{{#1}}${#2}}

\usepackage{draftwatermark}
\SetWatermarkText{\textsc{Draft}}
\SetWatermarkScale{6}

\newcommand{\be}{\begin{equation}}
\newcommand{\ee}{\end{equation}}

\newcommand{\bea}{\begin{eqnarray}}
\newcommand{\eea}{\end{eqnarray}}

\begin{document}

\title{Constraining level densities through quantitative correlations with cross-section data}

\author{G. P. A. Nobre}
\email[Corresponding author: ]{gnobre@bnl.gov}
\affiliation{National Nuclear Data Center, Brookhaven National Laboratory, Upton, NY 11973-5000, USA}

\author{D. A. Brown}
\affiliation{National Nuclear Data Center, Brookhaven National Laboratory, Upton, NY 11973-5000, USA}

\author{M. W. Herman}
\affiliation{National Nuclear Data Center, Brookhaven National Laboratory, Upton, NY 11973-5000, USA}
\affiliation{Los Alamos National Laboratory, Los Alamos, NM 87545, USA}

\author{A. Golas}
\affiliation{University of Massachusetts, Lowell, MA 01854, USA}

\begin{abstract}
The adopted level densities (LD) for the nuclei produced through different reaction mechanisms significantly impact the accurate calculation of cross sections for the different reaction channels.
Many common LD models make simplified assumptions regarding the overall behavior of the total LD and the intrinsic spin and parity distributions of the excited states. However, very few experimental constraints are taken into account in these models: LD at neutron separation energy coming from average spacings of $s$- and $p$-wave resonances ($D_0$ and $D_1$, respectively) whenever they have been previously measured, and the sometimes subjective extrapolation of discrete levels. These, however, constrain the LD only in 
very specific regions of excitation energy, and for specific spins and parities.
This work aims to establish additional experimental constraints on LD through  quantitative correlations between cross sections 
and LD. 
This allows for fitting and the determination of detailed structures in LD.
For this we use the microscopic Hartree-Fock-Bogoliubov (HFB) LD as a starting point as the HFB LD provide a more realistic spin and parity distributions than phenomenological models such as Gilbert-Cameron (GC).
We then associate variations predicted by the HFB model with the structure  observed in double-differential cross sections at low outgoing neutron energy, region that is dominated by the LD input. We also use ($n,p$) on \nuc{56}{Fe}, as an example case where angle-integrated cross sections are extremely sensitive to LD. For comparison purposes we also perform calculations with the GC model.
 With this approach we are able to perform fits of the LD based on actual experimental data, constraining the model and ensuring its consistency. This approach can be particularly useful in extrapolating the LD to nuclei for which high-excited discrete levels and/or values of $D_0$ or $D_1$ are unknown. It also predicts inelastic gamma ($n,n^{\prime}\gamma$) cross sections that in some cases can differ significantly from more standard phenomenological LD models such as GC.

\end{abstract}
\date{\today}

\maketitle

\setstcolor{red}

\section{Introduction}

As the excitation energy of a given nucleus increases, the number of excited states rises exponentially. Therefore, after a certain cutoff energy it becomes impractical to handle each level individually and one has to deal with the density of levels in order to properly determine the nuclear properties and associated cross sections. Several models have been proposed to describe the general behavior of level densities (LD), such as the Gilbert-Cameron \cite{GC}, Generalized Superfluid Model \cite{GSM-1,GSM-2}, Back-Shifted Fermi Gas~\cite{VonEgidy:1988,Dilg:1973}, or Enhanced Generalized Superfluid Model (EGSM) \cite{EGSM}. Those phenomenological models assume  simplified functional forms of the LD and their general behavior (spin and parity distributions, etc.), and they are constrained by limited availability of experimental data. 
For instance, resonance spacings, which are related to  the  LD at the neutron separation energy,   have only been experimentally measured for some nuclei, and they constrain the LD only at a single excitation energy point and only the LD for levels with specific spin and parity (this will be discussed in Section~\ref{sec:LD_at_Sn}). The other experimental constraint is at the intersection with measured discrete levels. Ideally, adopted LD should match the asymptotic behavior of the cumulative number of excited discrete levels. That, however, is often overlooked in nuclear data evaluations, favoring a LD parametrization that reproduces better an observed cross section at the expense of a realistic and smooth transition between discrete levels and level densities.

For a more  quantitative description of LD, many microscopic models have been  developed~\cite{HFBM,Nakada1997,Nakada1998,Alhassid1999,Zelevinsky2019,Senkov2013,Brown2019}, each adopting different approaches and approximations. Such microscopic models, having a more fundamental basis, tend to be more predictive in cases where little experimental information are available,  compared to phenomenological alternatives.  The  microscopic combinatorial Hartree-Fock-Bogoliubov (HFB) model \cite{HFBM} incorporated to the RIPL-3 parameter library \cite{RIPL-3} is an example of such models, offering a more global and self-consistent description of LD, even though in some cases it may not lead to agreement with data as good as the GC model. Also, while the GC model, like many phenomenological ones, simplistically assume  equal-parity and Gaussian-like spin distributions, the spin and parity distributions predicted by HFB are defined by  the arrangement of single-particle levels, although residual interaction is taken into account only approximately. Thus they are expected to be  
more reliable LD in the whole range of excitation energy, not only near the discrete-level cutoff or at the neutron separation energy $S_n$.

Quite often, nuclear reaction data evaluators  employ phenomenological LD models rather than microscopic ones due to the higher parameter-fitting flexibility of the former.  This can   lead to a better cross-section agreement with experiment (e.g., Ref.~\cite{WONDER2015}), at the expense of a more self-consistent description of the nuclear interaction. In this work, we  expand the work of Ref.~\cite{CNR*18}, showing how this apparent deficiency of the HFB LD model may be overcome by extracting experimental information from neutron double-differential spectra cross sections and other reaction channels in the case of neutron-induced reactions on  \nuc{56}{Fe}, and using this to impose constraints on the LD. The relationship between spectra and LD has been pointed out before~\cite{Voinov:2019}. Also, Ref.~\cite{Richter:1974} discusses the relationship between LD and cross sections, within the context of cross-section fluctuations. However, in our work we aim for establishing quantitative correlations within the context of complete reaction evaluations. Adopting the microscopic HFB model  leads to a more realistic and self-consistent description of the LD and cross sections that are in better agreement with experimental data when compared with the GC model,  in particular for the \nuc{56}{Fe}($n,p$) reaction which is both well-known and of interest for dosimetry  \cite{IRDFF1,IRDFF2}. We also obtain an improved description of inelastic-gamma cross sections from \nuc{56}{Fe}($n,n^{\prime}\gamma$) reaction allowing increased reliability for simulations of gamma transitions. 
This work represents a pathway to combine an accurate description of reaction observables with the predictive power of microscopic models, which will improve model calculations for many applications, such as astrophysics and radioactive-ion physics.

Additional constraints can be inferred in the future by the analysis of the experimental data recently obtained in the Oslo Cyclotron Laboratory \cite{Larsen2017} within the Oslo method~\cite{Schiller2000}. However,  to ensure a proper comparison, special care must be taken considering that the Oslo method makes model assumptions (e.g. assuming equal parity distribution) in order to disentangle LD and gamma strength function from the observable quantities actually measured. This has been discussed in Refs.~\cite{HFBM,Hilaire:2012}.

There are many other different LD models available in the literature (e.g. Shell-Model Monte Carlo \cite{Nakada1997,Nakada1998,Alhassid1999}, Moments-Method based Shell Model \cite{Zelevinsky2019,Senkov2013}, Extrapolated Lanczos Matrix~\cite{Brown2019}, etc.), each with their own advantages and simplifications. In this work we  restricted our analysis to the GC and HFB models, the former being a well-known, widely-accepted phenomenological model, while the latter is illustrative of a more fundamental, microscopic model. Both are representatives of their own class of models, and replacing either by another choice of phenomenological or microscopic model, while changing the details of calculations, would not be expected to substantially change the overall conclusions of the present work. Additionally, another reason for choosing the HFB model for LD as a representation of microscopic models in this work is the fact that it is the only one consistently available for the whole nuclear chart while others are not systematically applied.

\section{Background on LD models}
\label{sec:LD_models}

Phenomenological LD models tend to better reproduce average behaviors while missing detailed structure components. We will discuss the phenomenological Gilbert-Cameron and the microscopic HFB models, as they are defined in RIPL-3~\cite{RIPL-3} and implemented in the reaction code EMPIRE~\cite{empire,EmpireManual}. It is worth noting that, later in text, when we refer to Gilbert-Cameron calculations, we mean  the parametrization adopted  for the fast-region evaluation of \nuc{56}{Fe} present in the ENDF/B-VIII.0 \cite{CIELO-IRON,ENDF-VIII.0} as part of the CIELO project \cite{CIELO}, including cutoff energies where discrete levels transition to LD. Even though the RIPL-3 GC parametrization is based on global fits of GC parameters, which describe reasonably well cumulative level distributions and level spacings at $S_n$, that does not necessarily translate into good, consistent cross-section agreements in reaction calculations at the precision level required in evaluations. To optimize the agreement with cross-section data, the GC LD parameters were fitted in the ENDF/B-VIII.0 evaluation \cite{CIELO-IRON,ENDF-VIII.0}. The starting point for the HFB calculations will correspond to the same overall parametrization with the exception, of course, of the parameters related to the level densities. This ensures that the initial set of inputs lead to calculated cross sections that are  in good agreement with experimental data for all reactions.

\subsection{Gilbert-Cameron model}

Phenomenological LD models often assume at higher excitation energy some form of the analytical expressions of the Fermi Gas Model \cite{GC}. 
Assuming the approximation that the  density of intrinsic levels with spin $J$, parity $\pi$ and excitation energy $E_x$ can be factored  in terms of its excitation energy and spin and parity dependence:

\begin{equation}
\rho(E_x,J,\pi) = \tilde{\rho}(E_x) \hat{\rho}(J,\pi),
\end{equation}
where, for the Fermi-Gas model, we have  
\begin{equation}
\label{eq:GC-spindist}
\hat{\rho}^{\mathrm{FG}}(J,\pi)= \frac{2J+1}{2\sqrt{8\pi\sigma^{3}}} \mathrm{exp}\left[  - \frac{(J+1/2)^2}{2\sigma^2}\right],
\end{equation}
and
\begin{equation}
\label{eq:rho_FG_Ex}
\tilde{\rho}^{\mathrm{FG}}(E_x)= \frac{\pi}{12a^{1/4}U^{5/4}} \mathrm{exp}\left[  2\sqrt{aU} \right],
\end{equation}
where $\sigma^2$ is the spin cut-off parameter, $U$ is the effective energy ($U = E_x - \Delta$, where $\Delta$ is the pairing energy), and $a$ is the level-density parameter.

Within the Gilbert-Cameron model \cite{GC}, it is assumed that below a chosen matching excitation energy $U_x$ the LD can be described by a constant temperature formulation, given by:
\begin{equation}
\label{eq:rho_CT_Ex}
\tilde{\rho}^{\mathrm{CT}}(E_x)= \frac{1}{T} \mathrm{exp}\left[  \frac{E_x-E_0}{T}  \right], 
\end{equation}
where $T$ is the nuclear temperature and $E_0$ is a free parameter. Above $U_x$ the Fermi Gas excitation-energy component is given by Eq.~\ref{eq:rho_FG_Ex}, with pairing energy given by $\Delta = n\tfrac{12}{\sqrt{A}} $, where $A$ is the nucleus mass number and $n$ is 0, 1, or 2 for odd-odd, odd-even, and even-even nuclei, respectively. 
The parameter $U_x$ is 
internally determined by  imposing that the total LD and its derivative are continuous at the matching point $U_x$. The adopted values of $U_x$ in the GC calculations were the same as the ones in the ENDF/B-VIII.0 evaluation, namely 8.28 MeV for \nuc{56}{Fe} and 6.01 MeV for \nuc{56}{Mn}. The spin cut-off is given by $\sigma^2(E_x) = 0.146A^{2/3}\sqrt{aU} $.  One can use different systematics for the energy-dependency of the $a$ parameter in  Eqs.~\ref{eq:rho_FG_Ex} and~\ref{eq:rho_CT_Ex}. However, following original Gilbert-Cameron formulation,  constant $a$ were employed in the \nuc{56}{Fe} evaluation, fitted to reproduce experimental data.

\subsection{HFB model}
\label{sec:HFB-model}

There are many different formulations of the HFB model for nuclear LD~\cite{Hilaire:2001,Demetriou:2001,Hilaire:2006,HFBM}. In our present calculations we employed the microscopic combinatorial approach \cite{HFBM} documented in  RIPL-3 \cite{RIPL-3}, consisting of single-particle level schemes obtained from constrained axially symmetric Hartree-Fock-Bogoliubov method (HFBM) based on the BSk14 Skyrme force \cite{HFB-model} to construct incoherent particle-hole (ph) state densities $\omega_{\mathrm{ph}}(E_x,M,\pi)$ as functions of the excitation energy $E_x$, the spin projection $M$ (on the intrinsic symmetry axis of the nucleus) and the parity $\pi$.

Effects associated with collective degrees of freedom are taken into account  through the boson partition function as defined in Ref.~\cite{Hilaire2001}, which provides vibrational state densities dependent on multipolar phonon energies, while the shell corrections are the ones defined in Ref.~\cite{Goriely2002}. 
The adopted phonon energies, based on tabulated experimental vibrational levels, for quadrupole, octupole and hexadecapole phonons follow the ones established in Ref.~\cite{RIPL-3}.

\subsection{Spin and parity distributions}
\label{sec:spin-dist}

\begin{figure}[hptb]
\centering
\subfloat[Normalized spin distributions for levels with positive parity.]{ \label{fig:spin-dist-positive} \includegraphics[scale=0.70,keepaspectratio=true,clip=true,trim=0mm 0mm 0mm 0mm]{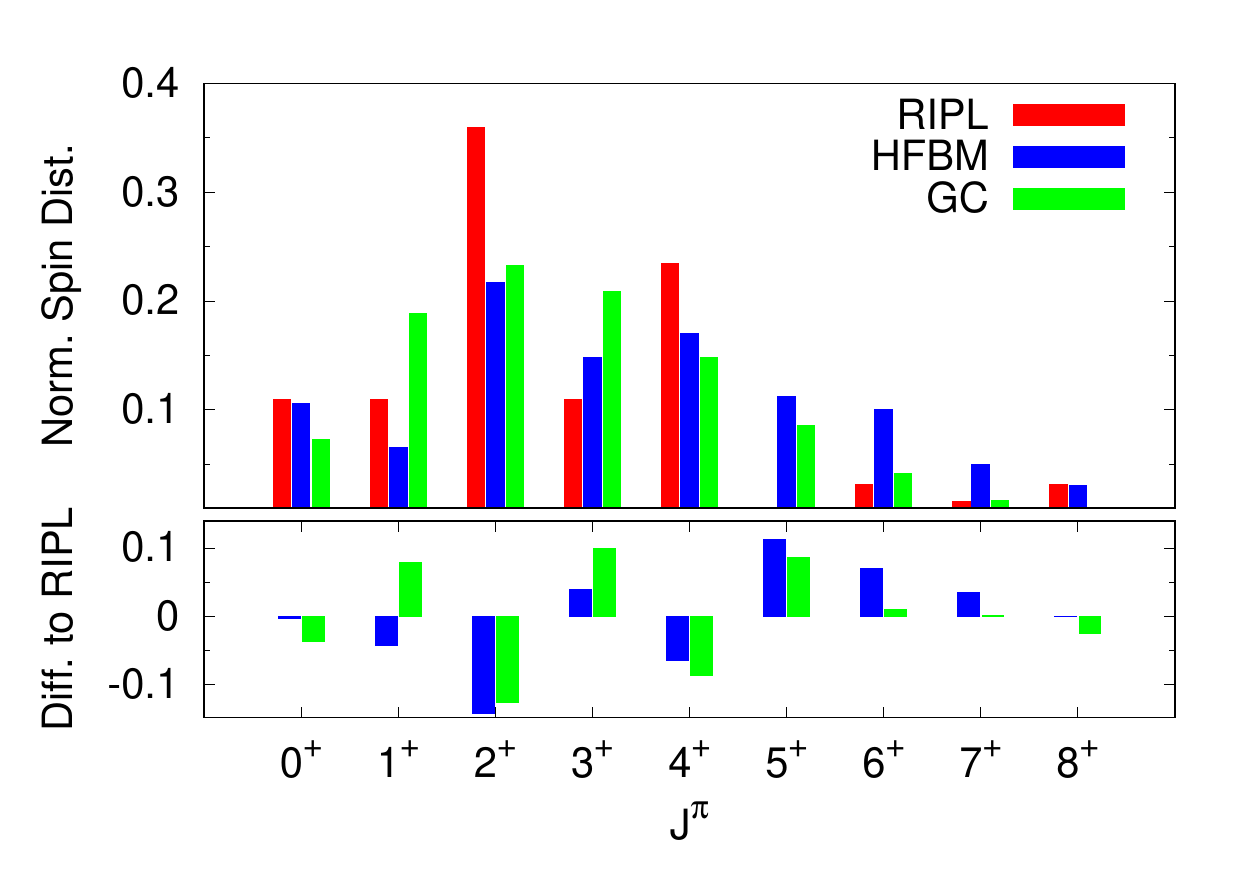}} \\
\subfloat[Normalized spin distributions for levels with negative parity.]{ \label{fig:spin-dist-negative} \includegraphics[scale=0.70,keepaspectratio=true,clip=true,trim=0mm 0mm 0mm 0mm]{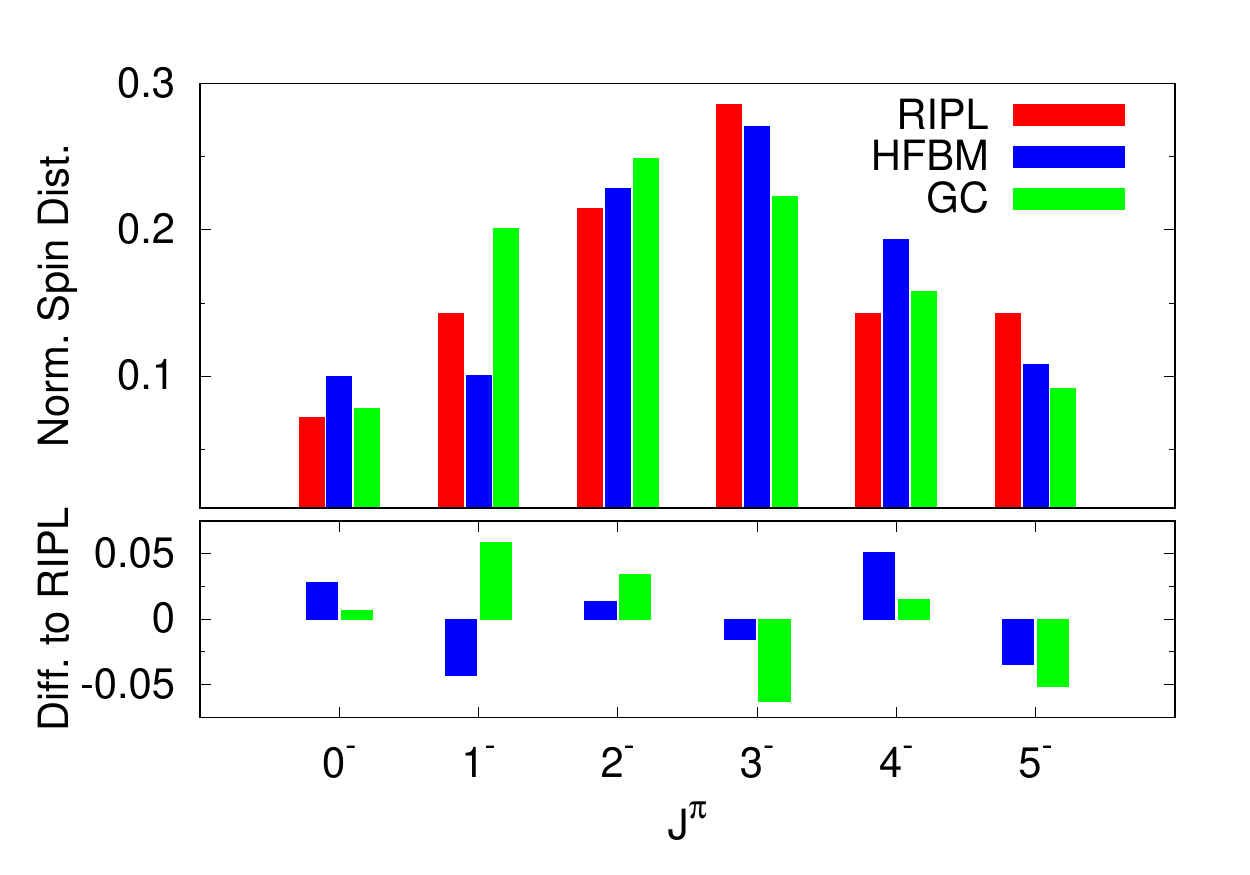}}
 \caption{Spin distributions for levels of positive (Figure~\ref{fig:spin-dist-positive}) and negative (Figure~\ref{fig:spin-dist-negative}) parities, up to the cut-off excitation energy of $E_{\mathrm{cut}}$ =  5.386~MeV. Results are shown for experimental discrete levels (as found in RIPL), and as predicted by the GC and HFB models. Each distribution is normalized by the total number of levels within each formalism. While the bottom panels show the difference between models and experiment.}\label{fig:spin-dist}
\end{figure}

\begin{figure}[h]
 \centering
 \includegraphics[scale=0.70,keepaspectratio=true,clip=true,trim=0mm 0mm 5mm 0mm]{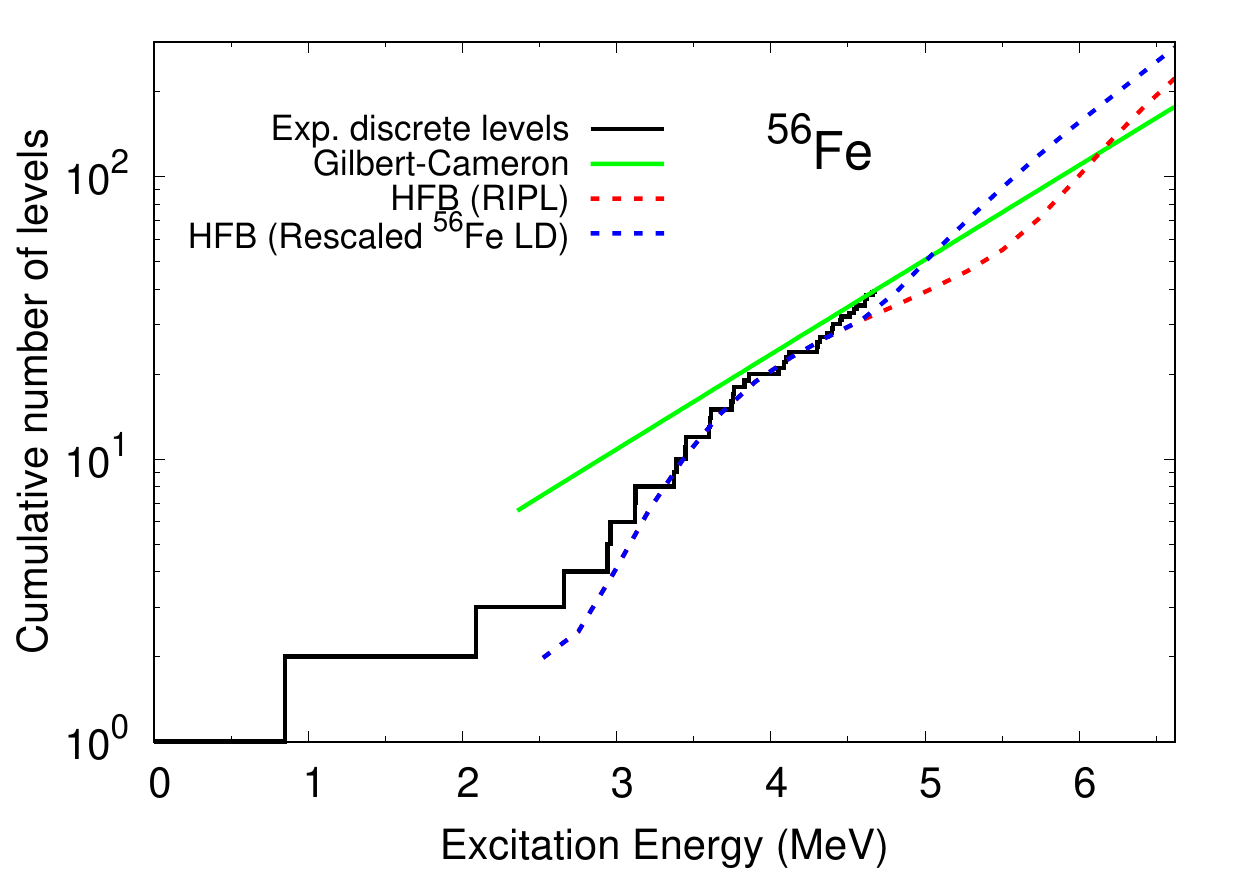}
 \caption{Cumulative number of levels for \nuc{56}{Fe}. The black curve is derived from the cumulative number of levels observed experimentally; the green curve is the LD from the GC model with its parameters fitted according to the ENDF/B-VIII.0 iron evaluation~\cite{CIELO-IRON}; the red dashed curve is the LD from the HFB model as tabulated in RIPL; the dashed blue curve is the HFB LD re-scaled to better agree with calculated neutron double-differential spectra.  See text for details on the calculations in each case.}
\label{fig:fe56-num_levels}
\end{figure}

We compared the distribution of the number of levels for each spin and parity from each model with what is experimentally observed, as stated in the levels segment of the RIPL library \cite{RIPL-3}. The red bars in Figure~\ref{fig:spin-dist} show the number of levels observed experimentally as contained in the RIPL library \cite{RIPL-3} for each spin and parity, normalized by the total number of levels for each parity, below a given energy $E_{\mathrm{cut}}$. This cut-off excitation energy was chosen to be $E_{\mathrm{cut}}$ =  5.386~MeV because above this excitation energy we begin to see levels with undetermined, or poorly-known, spins and/or parities in RIPL. In principle, by observing the experimental cumulative level distribution of \nuc{56}{Fe} (Figure~\ref{fig:fe56-num_levels}), we see that around 4 MeV there seem to be already some missing experimental levels, bringing down the derivative of the cumulative number of levels. 
However, due to the challenge of unambiguously defining the exact point at which observed levels are missing, we opted for the criterion above to define  $E_{\mathrm{cut}}$.

The total number of levels in RIPL with positive and negative parities were 64 and 14, respectively.  This asymmetry  is ignored within the GC model, and one of the consequences of such approximation will be discussed in Section~\ref{sec:inel-gammas}. The green bars in Figure~\ref{fig:spin-dist} show the spin distribution within the GC framework, which is the Gaussian distribution shown in Eq.~\ref{eq:GC-spindist} with the variance $\sigma$ = 2.591, again normalized so that the sum of the number of levels is 1, for each parity. The blue bars in Figure~\ref{fig:spin-dist} display the cumulative number of levels as a function of $J^{\pi}$, normalized to the total number of levels for each parity, as predicted according to the HFB model, by integrating the $J^{\pi}$-specific HFB LD up to $E_{\mathrm{cut}}$.

The observed spin distributions (RIPL) clearly do not follow a Gaussian distribution like GC does by construction.  This is the case for both positive (Figure~\ref{fig:spin-dist-positive}) and negative (Figure~\ref{fig:spin-dist-negative}) parities. The HFB ones, on the other hand are not  Gaussian  and clearly show structures, favoring one spin over the other. These structures in HFB spin distributions do not necessarily  match experimental observation. However, it is notable how well HFB describes the sharp decrease structure observed for 1$^+$ and 3$^+$ levels. To better visualize the different behaviors, in the bottom panels of both Figures~\ref{fig:spin-dist-positive} and~\ref{fig:spin-dist-negative} we show the difference between the normalized cumulative number of levels from both models relative to RIPL. It is important to note that, due to the adoption of a cutoff energy in the level counting, we introduce some uncertainty in the comparison with RIPL. Ideally, for the comparison between models and observed numbers of levels to be fair, all levels should be considered. In practical terms the cutoff in excitation energy should be very high, reducing the effects of the arbitrariety of the choice of $E_{\mathrm{cut}}$. For example, there are no observed 5$^+$ levels in \nuc{56}{Fe} below the chosen cutoff of $E_{\mathrm{cut}}$ =  5.386~MeV, but that does not mean that 5$^+$ levels would not to be expected at all above $E_{\mathrm{cut}}$. Likewise, counting levels with only one ($J^{\pi}$=7$^+$, 0$^-$) or two (6$^+$, 8$^+$,1$^-$,4$^-$,5$^-$) occurrences below $E_{\mathrm{cut}}$ are likely more dependent on the choice of cutoff.

\subsection{LD at the neutron separation energy}
\label{sec:LD_at_Sn}

The resonances observed in neutron-induced reactions on a given target nucleus are directly related to the excited-level scheme of the compound nucleus. The average spacing between $s$-wave resonances in the target nucleus, $D_0$, connects to the inverse of the level density in the compound nucleus at the neutron separation energy ($S_n$), for levels which with $J^\pi$ obtained from the coupling of the neutron spin and the ground state of the target nucleus. Analogously, a similar relation can be stablished for $p$-wave resonances ($L=1$).
Defining $\widetilde{S}_n = S_n+\Delta E/2$, where $\Delta E$ is the energy interval for which the resonances are determined (which is much smaller than $S_n$, so $\widetilde{S}_n\approx S_n$), this relation can be generalized in the following expression:
%
\begin{equation}
  \label{eq:DL-1}
    D_{L}^{-1} =  \sum_{J=J_{\textrm{min}}}^{J_{\textrm{max}}} \rho(\widetilde{S}_n, J , (-1)^L\pi_0) ,
\end{equation}
where $I_0$ and $\pi_0$ are respectively the spin and parity of the target nucleus,  $D_L$ is the average spacing of resonances of angular momentum $L$, and 
\begin{equation}
J_{\textrm{min}} = \textrm{max}(0,  |I_0 - L| - {\textstyle\frac{1}{2}})
\end{equation}
and
\begin{equation} 
J_{\textrm{max}} = I_0 + L + {\textstyle\frac{1}{2}}.
\end{equation}

The two cases of interest within this work are the LD for \nuc{56}{Fe} and \nuc{56}{Mn}, the former being the target nucleus and the latter is the residual of the ($n,p$) reaction. Information about such LD at $E_x=S_n$ should be then obtained from the resonance spacings of neutron-induced reactions on the target nuclei \nuc{55}{Fe} and \nuc{55}{Mn}, respectively. Even though Ref.~\cite{Atlas2018-vol1} provides both $D_0$ and $D_1$ for \nuc{55}{Mn}, there are no experimental values for \nuc{55}{Fe} as it is not a stable nucleus. For this reason, in the following discussion we focus on the \nuc{56}{Mn} LD at $S_n$. Approaches such as interpolation or systematics could in principle provide values of $D_0$ and/or $D_1$ for \nuc{55}{Fe}. However, the focus of the present work  is on direct experimental constraints on LD. 

\begin{figure}[h]
 \centering
 \includegraphics[scale=0.35,keepaspectratio=true,clip=true,trim=0mm 13mm 4mm 25mm]{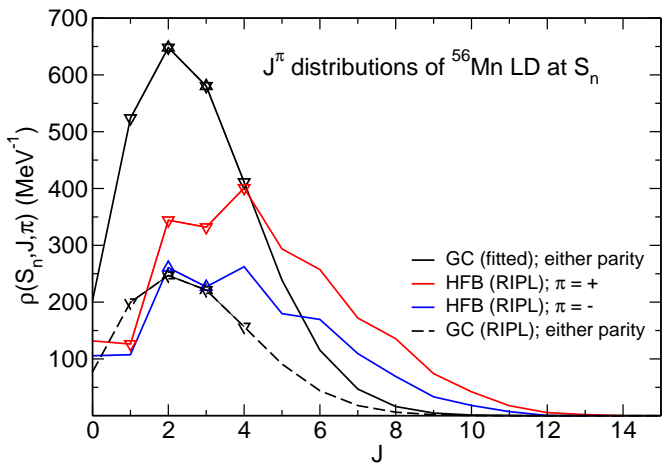}
 \caption{Spin and parity distributions of the \nuc{56}{Mn} level density at the neutron separation energy ($E_x=S_n$). Black solid curve corresponds to the GC model with the ENDF/B-VIII.0 parametrization for either parity; red and blue curves correspond to the HFB model as defined in RIPL-3~\cite{RIPL-3} for positive and negative parities, respectively; black dashed corresponds to GC model but with parametrization from RIPL-3. Upside and downside triangles highlight the spins and parities that contribute to $D_0^{-1}$ and $D_1^{-1}$, respectively.}
\label{fig:mn56-LD_at_Sn}
\end{figure}

Figure~\ref{fig:mn56-LD_at_Sn} shows the spin distributions of \nuc{56}{Mn} LD at the neutron separation energy ($S_n=7.27044 \pm 0.00013$ MeV) for the GC  and HFB models. In the case of GC, the solid black curve represents the LD obtained with  the parametrization used in the ENDF/B-VIII.0 \nuc{56}{Fe} evaluation, i.e.\ the parametrization that best reproduced \nuc{56}{Fe}($n,p$)\nuc{56}{Mn} cross sections. For comparison purposes, the black dashed line represents the GC with parametrization from RIPL. The spin and parity distributions from the HFB model (as parametrized in RIPL) are represented by the red (positive parity) and blue (negative parity) curves. Due to the equal-parity distribution assumption in the GC model, the black curves represent either parity.

 Also, considering the $\sfrac{5}{2}^{-}$ ground state of \nuc{55}{Mn}, we show in Figure~\ref{fig:mn56-LD_at_Sn}  as upside triangles the spin/parities that contribute to $D_0$ ($J^{\pi}=2^-,3^-$) and as downside triangles the ones contributing to $D_1$ ($J^{\pi}=1^{+},2^{+},3^{+},4^{+}$), following Eq.~\ref{eq:DL-1}. From this we calculate the  $D_0^{-1}$ and $D_1^{-1}$ values obtained from the different approaches for \nuc{56}{Mn} LD and compare with the experimental values from Ref.~\cite{Atlas2018-vol1}\footnote{The level spacings in \nuc{56}{Mn} correspond to the resonance  spacing of neutron-induced reactions on \nuc{55}{Mn}.}. We present these in Table~\ref{tab:D0_D1}.

\begin{table}
\caption{Comparison between the experimental  $D_0^{-1}$ and $D_1^{-1}$, in units of MeV$^{-1}$ 
with values obtained from \nuc{56}{Mn} level density.}
\label{tab:D0_D1}
 \begin{tabular}{l|c|ccc}
  \hline 
  \hline
                      & exp.~\cite{Atlas2018-vol1}     & GC (fit) & HFB (RIPL) &GC (RIPL) \\

\cline{2-5}
$D_0^{-1}$   &  413 $\pm$ 25   & 1228   & 488              & 468 \\
$D_1^{-1}$   &  909 $\pm$ 83   & 2163   & 1203           & 824 \\
  \hline
  \hline
 \end{tabular}
\end{table}

We can notice that LD of few spins and specific parity contribute to $D_0$ or $D_1$ and, due to different model-assumptions of spin and parity distributions,  similar calculated $D_0^{-1}$ and $D_1^{-1}$ can lead to very different total LD at neutron separation energy. Therefore, relying solely on resonance spacings to normalize total LD significantly limits the accuracy of the experimental constraint imposed onto the LD. 
We also  draw attention to the fact that, by comparing  the two GC approaches, we note that in order to  obtain optimal cross-section agreement, the agreement with resonance-spacing measurements is destroyed, leading to an  inconsistency between LD and cross section description.
Another noteworthy aspect is that, even at relatively high excitation energies, microscopic LD models predict non-equal parity distributions and ``non-Fermi-Gas'' spin distributions. Therefore making those assumptions when calculating total LD from resonance spacing introduces often-unquantified uncertainties to the final values. 


\section{Impact of LD models in cross sections}
\label{sec:xsec}

\begin{figure}[h]
 \centering
 \includegraphics[scale=0.70,keepaspectratio=true,clip=true,trim=0mm 0mm 5mm 0mm]{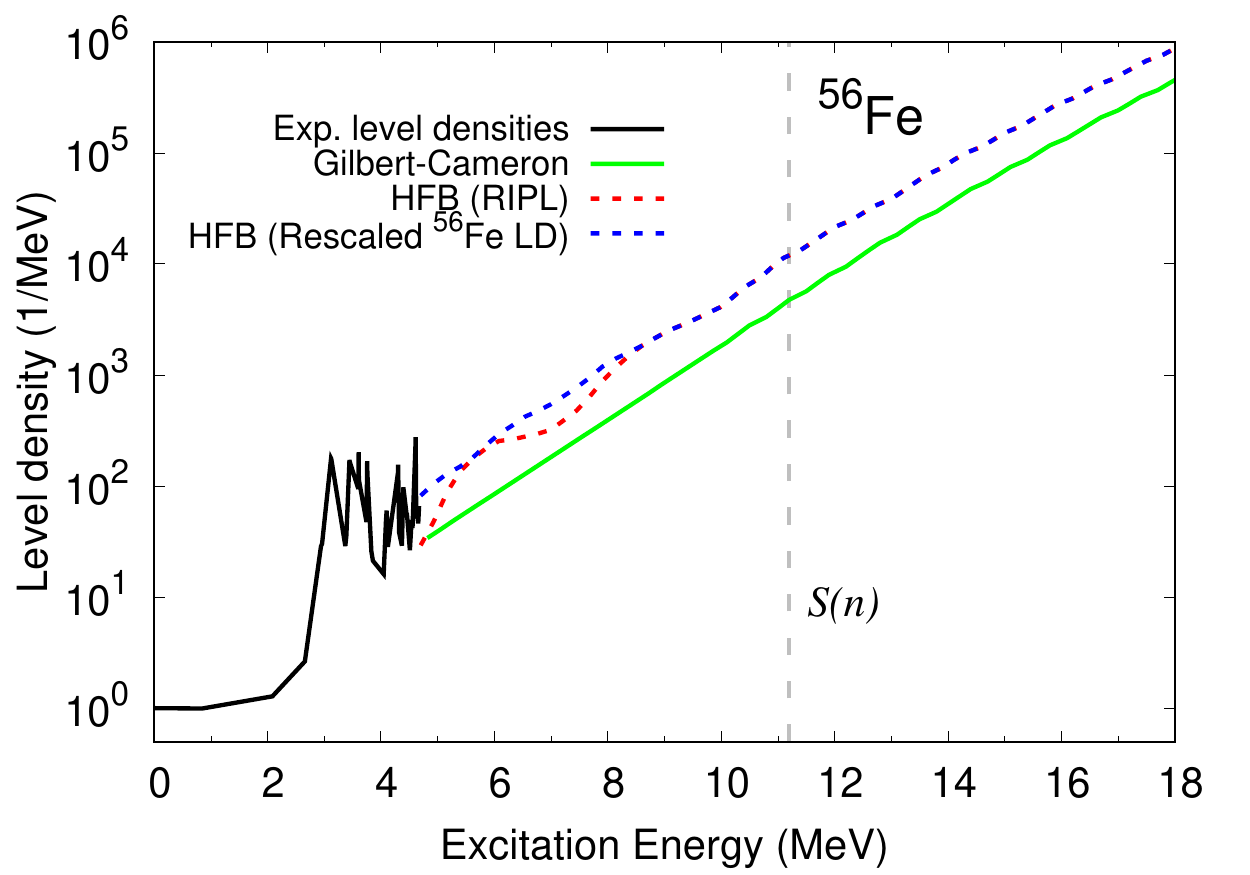}
 \caption{Level densities for \nuc{56}{Fe}. The meaning of the curves is the same as of Figure~\ref{fig:fe56-num_levels}. The dashed gray line marks the neutron separation energy  $S_n$  of \nuc{56}{Fe}.
 }
 \label{fig:fe56-ld}
\end{figure}
\begin{figure}[h]
 \centering
 \includegraphics[scale=0.72,keepaspectratio=true,clip=true,trim=0mm 0mm 4mm 0mm]{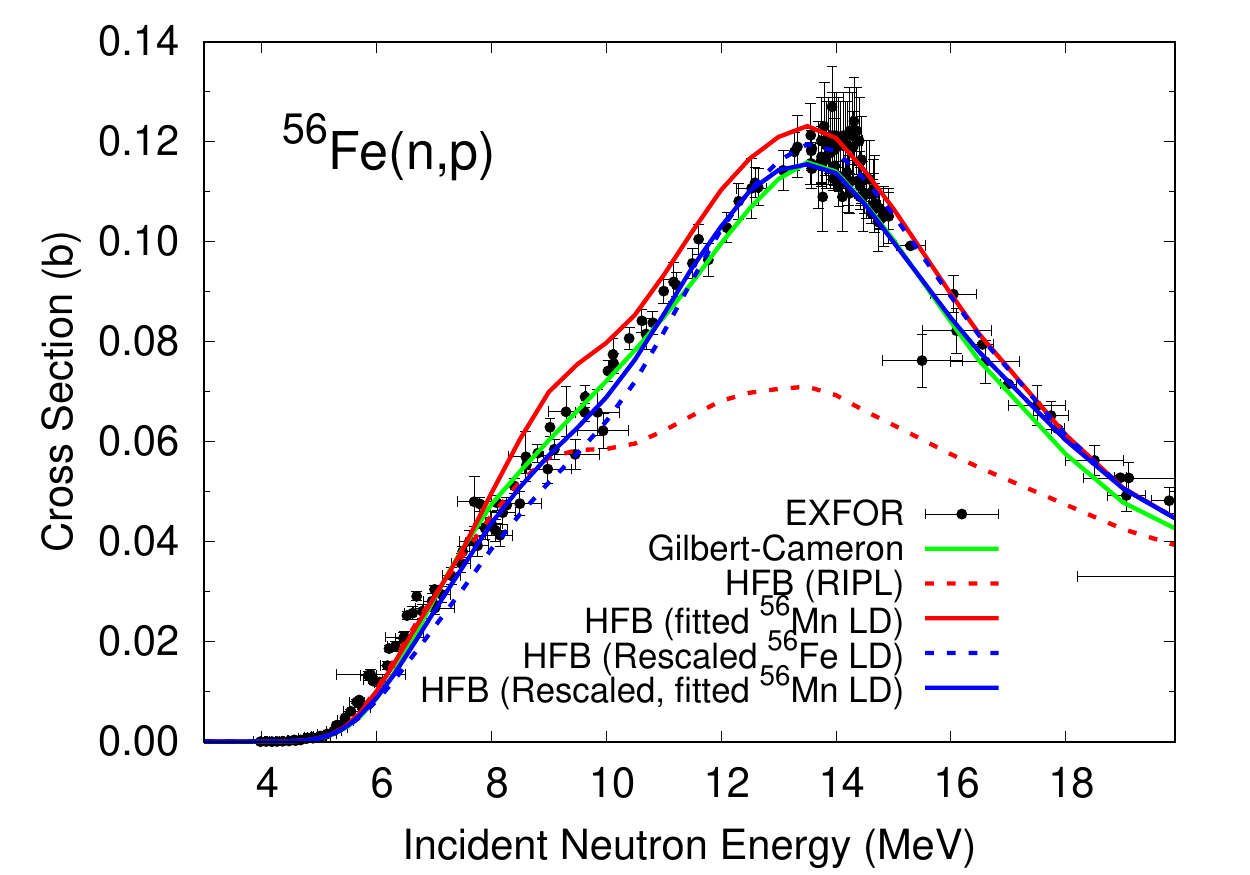}
 \caption{Cross section for the \nuc{56}{Fe}(n,p)\nuc{56}{Mn} reaction calculated using the different assumptions for the LD, as detailed in text. The green curve is the LD from the GC model with its parameters fitted according to the ENDF/B-VIII.0 iron evaluation~\cite{CIELO-IRON}; the red dashed curve is the LD from the HFB model as tabulated in RIPL; the red solid curve is the result after fitting \nuc{56}{Mn} HFB LD parameters to optimize agreement with ($n$,$p$) data; the blue dashed curve is the same as solid-red but also with HFB LD for \nuc{56}{Fe} re-scaled to better agree with calculated neutron double-differential spectra; the solid blue curve is the same as previous but also with \nuc{56}{Mn} re-scaled and re-fitted to ($n$,$p$) data. See text for details on the calculations of each curve. Experimental data retrieved from EXFOR~\cite{EXFOR,EXFOR-2}.
 }
 \label{fig:fe56-np}
\end{figure}

As our starting point to investigate the impact and correlations of details of LD in the cross sections, we adopted the parametrization employed in the  ENDF/B-VIII.0 evaluation for  \nuc{56}{Fe} in neutron-induced-reactions \cite{ENDF-VIII.0,CIELO-IRON}. This allowed us to begin the calculations with  a  set of parametrizations that produce  consistent differential and angle-integrated cross sections for all relevant reactions that are in good agreement with experimental data.  We can directly compare the total LD from both GC and HFB models, as seen on Figure~\ref{fig:fe56-ld}. 
The green curve in Figure~\ref{fig:fe56-ld} corresponds to the Gilbert-Cameron model for the LD of all nuclei, with parameters fitted to optimize the overall agreement with experimental data, as explained in Section~\ref{sec:LD_models}. 
The red dashed curve in Figure~\ref{fig:fe56-np} is the result of the same calculation but replacing the LD model by the  HFB one  described in Section~\ref{sec:HFB-model} and taken from RIPL-3 \cite{RIPL-3}, without any modifications. 

We  see that, even though the LD are approximately the same as the LD at the matching point  from experimental discrete levels (Figure~\ref{fig:fe56-ld}), they differ in the asymptotic behavior for high excitation energies $E_x$. Also important is the fact that, while the Gilbert-Cameron LD is smooth (as it comes from the constant-temperature analytical forms in Eq.~\ref{eq:rho_CT_Ex}), the HFB LD  fluctuates in the range  $5 \lesssim E_x \lesssim 9$~MeV. Figure~\ref{fig:fe56-num_levels} shows the cumulative number of levels for the different calculations using the same choice of colors for the curves. Both Gilbert-Cameron and HFB (from RIPL) models approximately reproduce  the number of levels at around 4.5~MeV which is around where one would normally impose the transition from the discrete levels to LD. This transition point, or excitation energy cut-off, can  be rather arbitrary. In this case of \nuc{56}{Fe}, it seems that any value between $\approx$ 3.7 and $\approx$ 4.5 MeV should be an equally good choice  for the cut-off, but this may not be the case for other nuclei. One can clearly see from Figure~\ref{fig:fe56-num_levels} that the HFB predicted cumulative number of levels is in a much better agreement with the behavior of observed discrete levels, which makes it more independent of  the choice of excitation energy at which the transition to the LD is made.

We initially compare the performance of both LD models when applied to \nuc{56}{Fe} by observing their impact on  \nuc{56}{Fe}($n$,$p$), which is a well-measured  dosimetry reaction \cite{IRDFF1,IRDFF2}. 
In the incident-energy region where ($n$,$p$) is prominent, it is the only relevant open channel apart from elastic and inelastic channels, which are much bigger to be significantly impacted by details of ($n$,$p$) and by fine changes in LD \cite{CIELO-IRON}. The ($n,2n$) channel only opens above $\approx$ 11.5 MeV.
Neutron capture is obviously open, but its cross section is orders of magnitude smaller than ($n$,$p$), making the latter the ideal mechanism to probe the LD associated with \nuc{56}{Fe} and \nuc{56}{Mn}. 

In Figure~\ref{fig:fe56-np} we present results for the \nuc{56}{Fe}($n$,$p$)\nuc{56}{Mn} cross sections from different calculations employing different approaches for the LD. The colors of the curves represent the same calculations as in Figures~\ref{fig:fe56-num_levels} and~\ref{fig:fe56-ld}, namely green for fitted Gilbert-Cameron and dashed-red for default HFB model, while the other curves in Figure~\ref{fig:fe56-np} will be explained later in the text.  Clearly, blindly using the HFB LD as they are provided in RIPL-3 produces in this case a very poor agreement with experimental data. It is important to mention that RIPL provides correction tables for the HFB LD, taken from Ref.~ \cite{HFBM}, which aim to improve the overall agreements with experimental discrete level sequences and $D_0$. Such corrections were always taken into account in our calculations, whether the raw HFB LD had been rescaled or not before the corrections were applied.

\begin{figure}[h]
 \centering
 \includegraphics[scale=0.70,keepaspectratio=true,clip=true,trim=0mm 0mm 5mm 0mm]{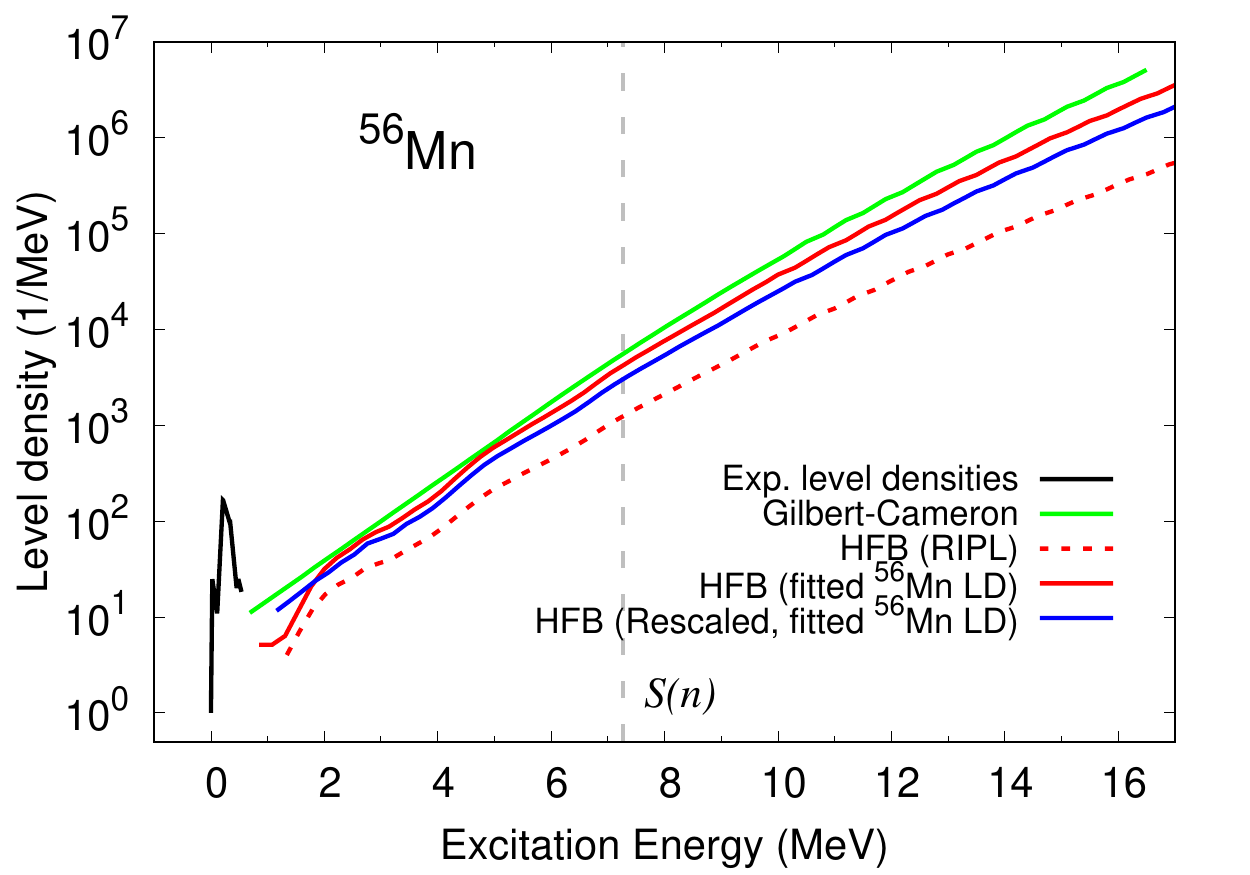}
 \caption{Level densities for \nuc{56}{Mn}. The black curve is derived from the cumulative number of observed experimentally; the green curve is the LD from the GC model with its parameters fitted according to the ENDF/B-VIII.0 iron evaluation~\cite{CIELO-IRON}; the red dashed curve is the LD from the HFB model as tabulated in RIPL; the red solid curve is the LD after fitting \nuc{56}{Mn} HFB LD parameters to optimize agreement with ($n,p$) data; the blue solid curve is the \nuc{56}{Mn} HFB LD re-scaled and re-fitted to ($n,p$) data after having re-scaled  HFB LD for \nuc{56}{Fe} to better agree with calculated neutron double-differential spectra. The dashed gray line marks the neutron separation energy $S_n$ of \nuc{56}{Mn}. 
See text for details on the calculations of each curve.} 
 \label{fig:mn56-ld}
\end{figure}

\begin{figure}[h]
 \centering
 \includegraphics[scale=0.70,keepaspectratio=true,clip=true,trim=0mm 0mm 5mm 0mm]{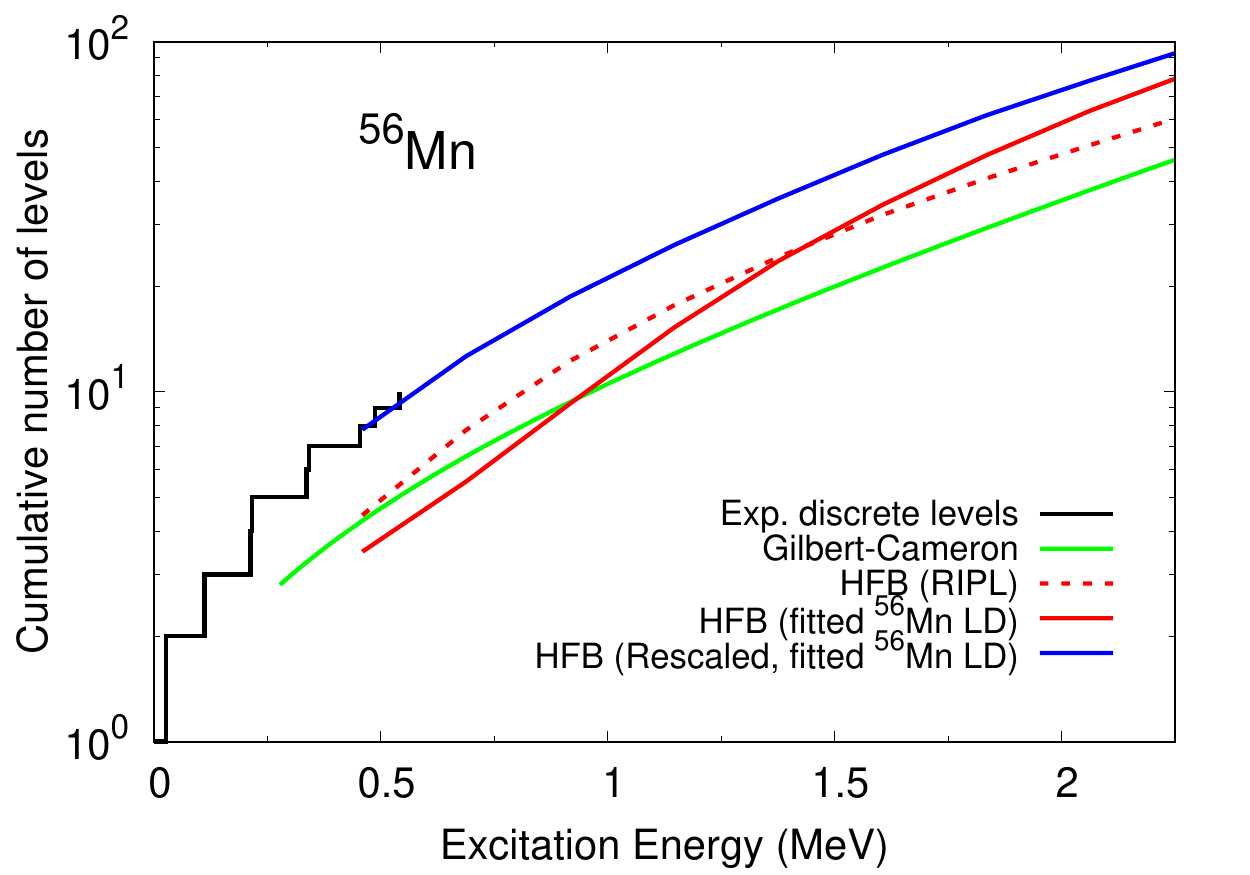}
 \caption{Cumulative number of levels for \nuc{56}{Mn}. The meaning of the curves is the same as of Figure~\ref{fig:mn56-ld}.}\label{fig:mn56-num_levels}
\end{figure}

One can rightly claim that the comparison with the Gilbert-Cameron result is not  fair since the calculation with Gilbert-Cameron had gone through parameter fitting. With this in mind we used the fitting code KALMAN \cite{KALMAN} within the EMPIRE package \cite{empire,EmpireManual} to vary the two parameters associated with the \nuc{56}{Mn} HFB LD, finding values which minimized the $\chi^2$ of calculated cross sections in relation to experimental data for all relevant reactions. Within EMPIRE, 
those parameters 
are basically scaling of parameters related to $a$ from Eqs.~\ref{eq:rho_FG_Ex} and~\ref{eq:rho_CT_Ex}  and of the excitation-energy shift.
After the fit, the optimal parametrization found was to increase 
one of the parameters by 45\% and the other one by 49\%. This is analogous to the procedure performed in the \nuc{54,56,57,58}{Fe} evaluations  \cite{CIELO-IRON} where LD parameters, in those cases corresponding to the Gilbert-Cameron model, were fitted to reproduce observed cross sections.  The effect of such fits of \nuc{56}{Mn} HFB LD parameters can be seen in Figures~\ref{fig:mn56-ld} and~\ref{fig:mn56-num_levels} as the red solid curves (again, green curves correspond to Gilbert-Cameron LD model and dashed-red to HFB model as in RIPL-3).

The result of such calculations with fitted \nuc{56}{Mn} LD is represented by the solid-red curve in Figure~\ref{fig:fe56-np}. We can see that, even though the fit leads to an improvement in the \nuc{56}{Fe}($n$,$p$) cross section (solid-red curve compared to dashed-red one), the agreement with experimental data is still not as good as the one from Gilbert-Cameron model (green curve). However, the improvement in agreement with the ($n$,$p$) cross-section data did not mean that the \nuc{56}{Mn} LD is indeed better than the unfitted (RIPL) one, as both fail to match the observed discrete levels, as seen in Figure~\ref{fig:mn56-num_levels} (solid-red and dashed-red curves, respectively). As a matter of fact, even the Gilbert-Cameron calculation, which reproduces well the \nuc{56}{Fe}($n$,$p$), uses \nuc{56}{Mn} LD which does not agree well with observed discrete levels (green curves on both Figures~\ref{fig:mn56-num_levels} and~\ref{fig:fe56-np}). This indicates that a better cross section agreement does not necessarily imply that a more realistic LD was employed. Ideally a realistic model for LD  should be able to consistently describe discrete levels, $D_0$ when available, as well as angle-integrated and differential cross sections.

\subsection{Relation between spectra and LD}
\label{sec:spectra_and_LD}

\begin{figure*}[htb]
 \centering
 \includegraphics[scale=0.70,keepaspectratio=true,clip=true,trim=0mm 5mm 5mm 0mm]{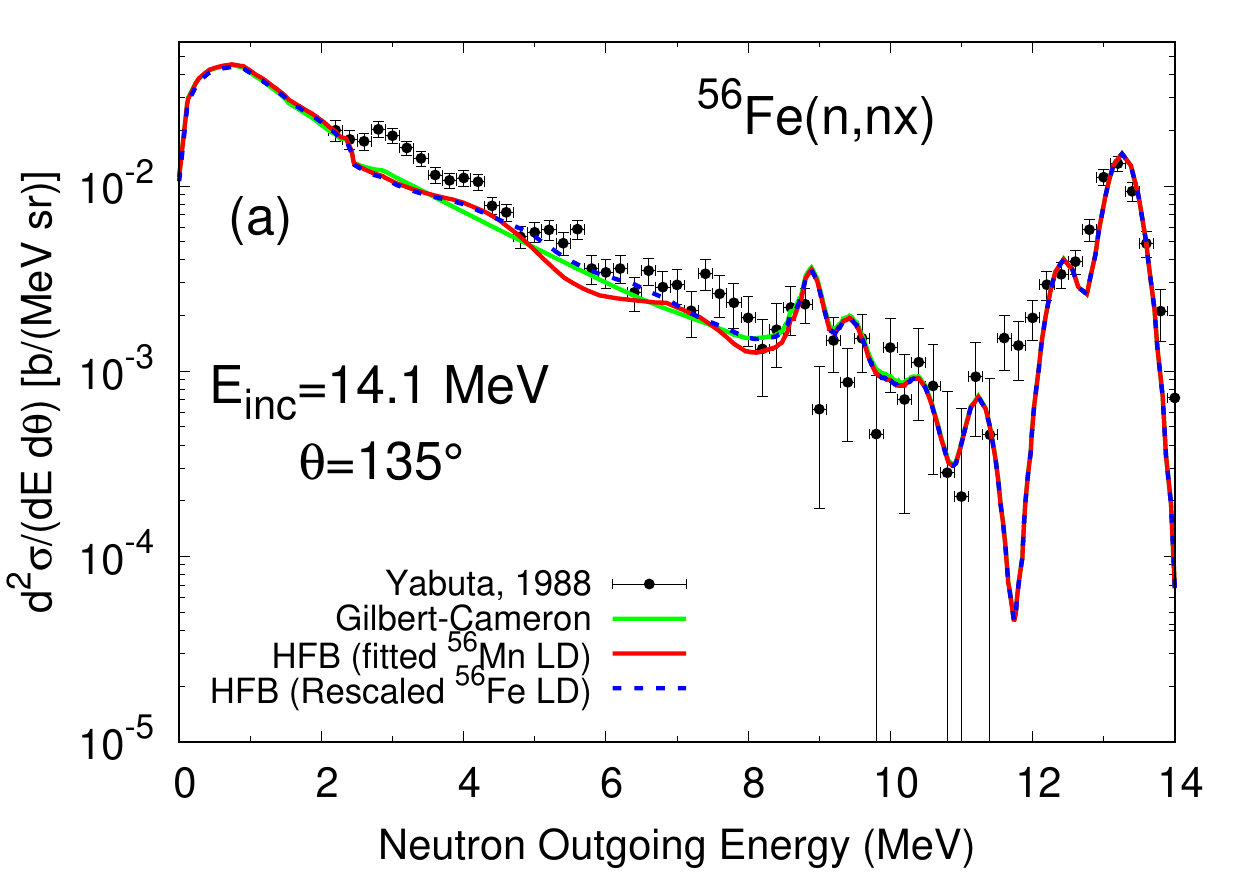} 
 \includegraphics[scale=0.70,keepaspectratio=true,clip=true,trim=0mm 5mm 5mm 0mm]{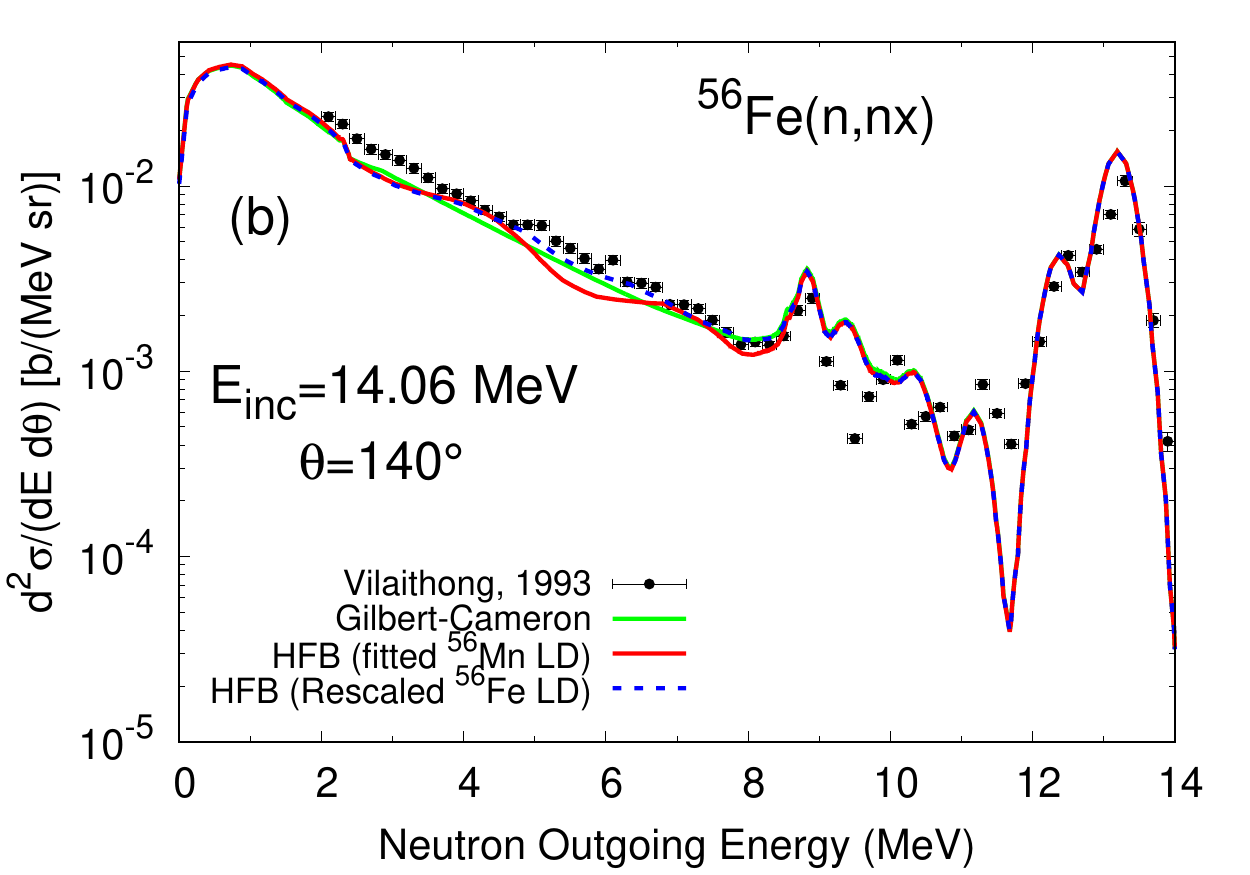} \\
 \includegraphics[scale=0.70,keepaspectratio=true,clip=true,trim=0mm 0mm 5mm 3mm]{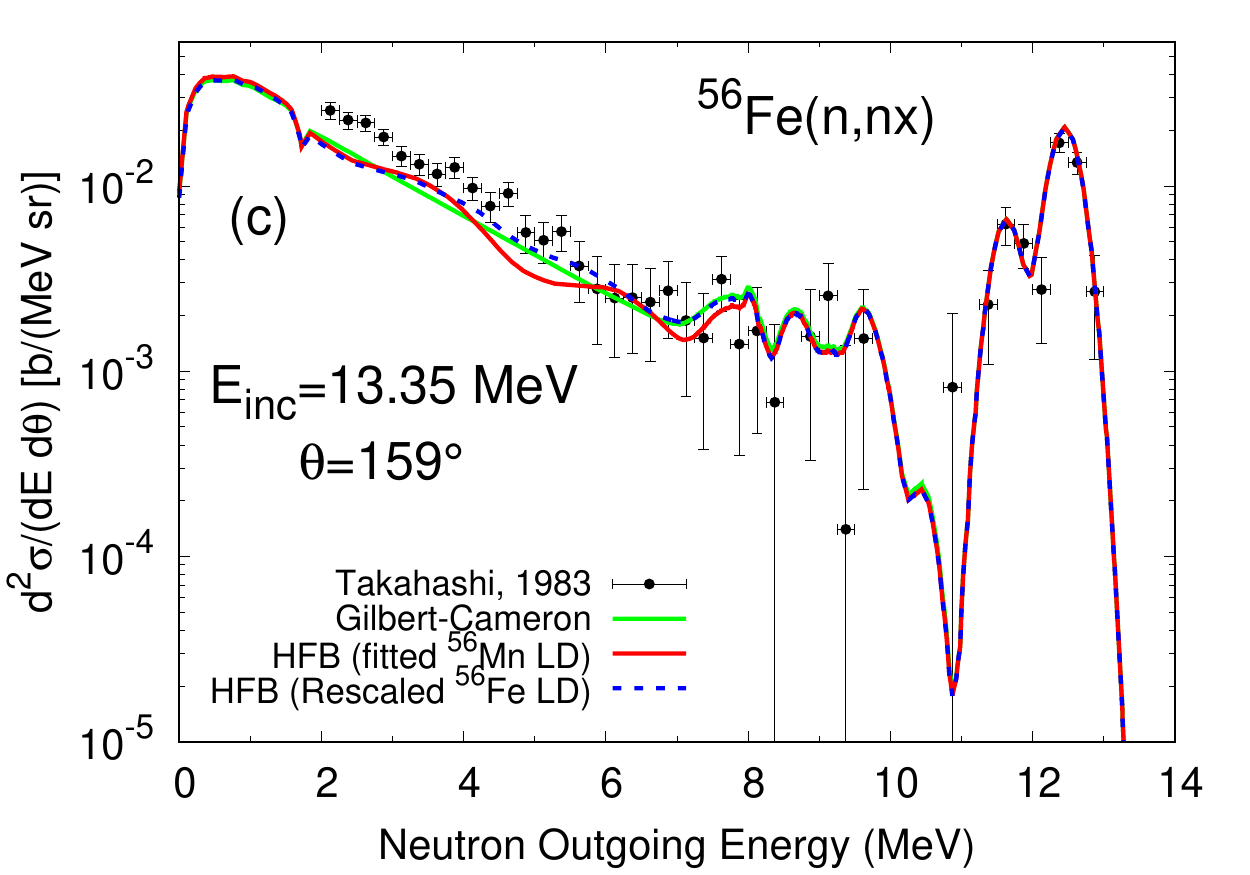} 
 \includegraphics[scale=0.70,keepaspectratio=true,clip=true,trim=0mm 0mm 5mm 3mm]{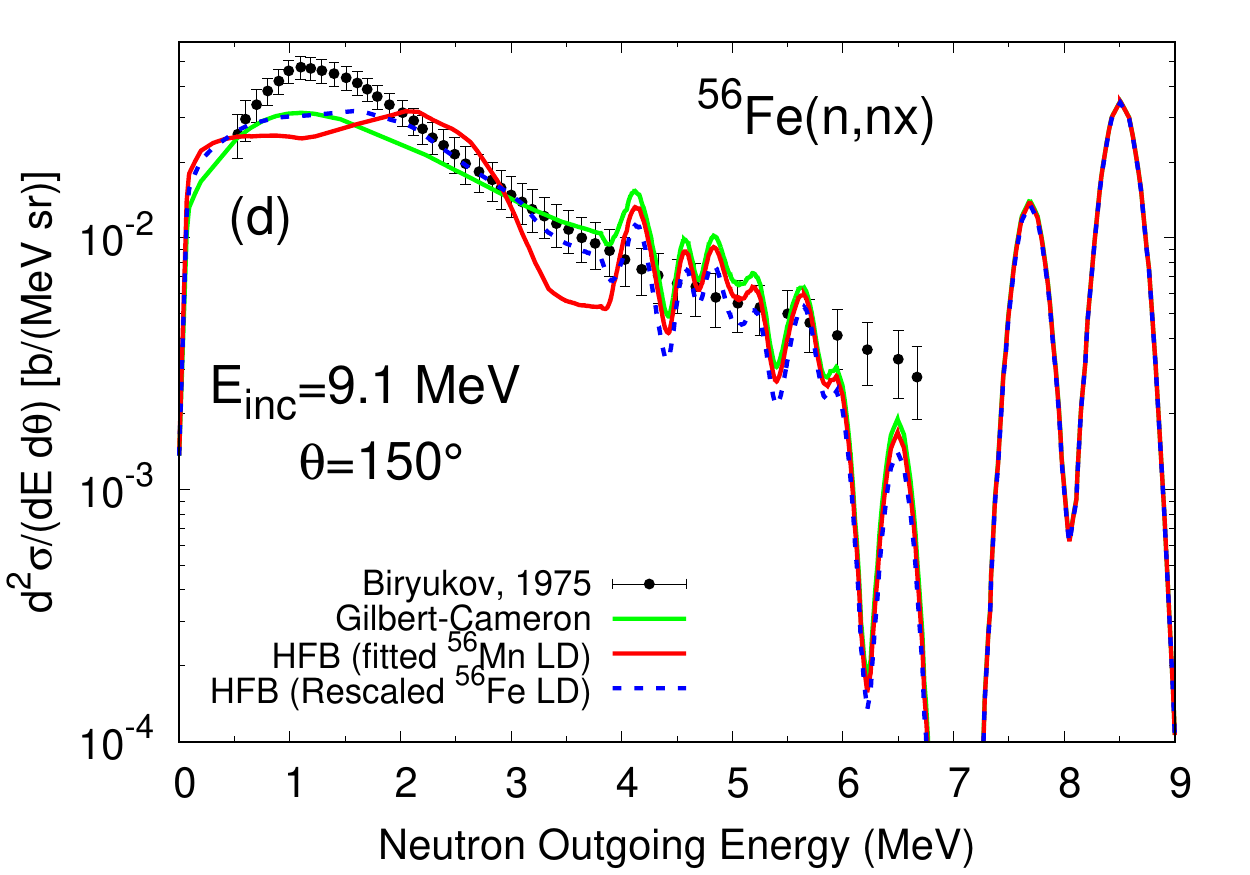} \\
 \caption{Example of double-differential spectra for different neutron incident energies and at different scattering angles for the different LD approaches. The green curves are the results from using the  LD from the GC model with its parameters fitted according to the ENDF/B-VIII.0 iron evaluation~\cite{CIELO-IRON}; the solid-red curves are the results from  using  the HFB model and  fitting \nuc{56}{Mn} HFB LD parameters to optimize agreement with ($n,p$) data; the dashed-blue curve is the same as the solid-red one but also with HFB LD for \nuc{56}{Fe} re-scaled to better agree with calculated neutron double-differential spectra. Data retrieved from EXFOR \cite{EXFOR,EXFOR-2}.
 }\label{fig:ddx}
\end{figure*}

The main purpose of this Section is to develop a set of prescriptions to adapt the HFB model to address its limits as presented above and in Section \ref{sec:LD_models}, providing cross sections as reliable as the ones obtained from the phenomenological Gilbert-Cameron LD model.  To this end, we also investigated  the impact of different LD models on  the behavior of neutron double-differential spectra. In Figure~\ref{fig:ddx} we can see that while the Gilbert-Cameron calculation (green curve) is in reasonable agreement with experimental data, the HFB one (red solid curve) has oscillations in the lower neutron-outgoing energy ($E_\mathrm{out}$) region that are not seen in data. This can be seen at around 3 MeV $<$ $E_\mathrm{out}$ $<$ 7 MeV for the incident energies of $E_\mathrm{inc}$=14.1, 14.06, and 13.35 MeV; and 1 MeV $<$ $E_\mathrm{out}$ $<$ 3 MeV for  $E_\mathrm{inc}$=9.1 MeV. 

We note that the  oscillations seen in the double-differential (DD) neutron spectra (Figure~\ref{fig:ddx}) have a direct correspondence to the structures observed in the \nuc{56}{Fe} HFB LD (Figure~\ref{fig:fe56-ld}, red dashed curve). Therefore, we performed a pointwise re-scaling of the  \nuc{56}{Fe}  HFB LD in the excitation energy ($E_x$) region below around 8 MeV in order to reduce the oscillations in the DD spectra and improve its agreement with data. 
This re-scaling of HFB LD was performed by simply multiplying  each tabulated value of the LD by an excitation-energy-dependent factor, iteratively, so that the agreement with spectra data obtained by the corresponding calculation would be gradually improved. Even though this procedure may be regarded as somewhat \emph{ad hoc}, this was a proof of principle that we can establish a quantifiable direct correlation between details and structures of cross-section spectra and LD, using the former to constrain the latter. Even though it seems to be an arbitrary modification, it actually leads to smoothing of the HFB fluctuating structure. The effects of missing residual interactions in HFB LD were originally simulated  via energy-broadening (smoothing) of the fluctuations resulting from combinatorial  calculations. Additional smoothing required in the present work  might indicate that the original smoothing should be more aggressive to better account for the residual interactions. A similar effect has been observed in \nuc{43}{Sc}~\cite{Burger:2012}, which was attributed to particle-vibration coupling not properly accounted for in the HFB model.
A  satisfactory agreement with the DD data, obtained with the rescaled HFB LD, is shown by the dashed-blue curves in Figure~\ref{fig:ddx}. This rescaled LD and the corresponding cumulative number of levels are shown as the dashed-blue curves in Figures~\ref{fig:fe56-ld} and~\ref{fig:fe56-num_levels}, respectively.

The  \nuc{56}{Fe} LD as presented in Figure~\ref{fig:fe56-ld} is not necessarily the optimal one. Rather we establish that we can use experimental data from double-differential measurements to impose constraints in the level densities in excitation-energy regions where no direct experimental information is available. This should improve the overall consistency between the LD for the different nuclei and also improve the model self-consistency for the calculated cross sections. As a matter of fact, if there were  sufficiently well-measured DD neutron spectra so that to confirm the existence of certain structures in the pre-equilibrium region of the neutron spectra, these same structures could be likely reproduced by imposing fluctuations in the LD.

It is  possible that the  smoothing of naturally-occurring  fluctuations in the combinatorial calculations was insufficient. Such smoothing simulates the effect of the residual interactions missing in the calculations, which in turn correspond to the underestimation of the effect of residual interactions. 
This indicates that it is  possible to use such reaction data-based constraints to improve 
the development of microscopic LD models, leading to more realistic predictions.


By rescaling the  \nuc{56}{Fe}  HFB LD to improve the neutron DD spectra, we also improve  the calculated ($n$,$p$) cross section, as  can be seen as the blue dashed curve in Figure~\ref{fig:fe56-np}. However, this agreement does not seem to be as good as the one obtained by the Gilbert-Cameron LD (green curve). With this in mind, we decided to also smooth the structures in the \nuc{56}{Mn} LD  and perform a new fit of their corresponding HFB parameters. 
The result  is shown as the solid-blue curves on Figures~\ref{fig:fe56-np},~\ref{fig:mn56-ld}, and~\ref{fig:mn56-num_levels}. We can see in Figure~\ref{fig:fe56-np} that now the calculated cross section is in an equally-good agreement with experimental data when compared to the Gilbert-Cameron calculation. 
One could even say that, except in  the region between 8 and 10 MeV where GC is better,  the new calculation  agrees with experimental data as well or  better than Gilbert-Cameron. As a self-consistency byproduct of this approach, the calculated final level densities and the related cumulative number of levels are in better agreement with observed levels than Gilbert-Cameron, as it can be seen when comparing the solid-blue and green curves in Figure~\ref{fig:mn56-num_levels}. 

It is noteworthy that the final values of the LD parameters in EMPIRE 
were in this fit raised from the default configuration only 33\% and 18\% in comparison with 45\% and 49\%, respectively, as stated in Section~\ref{sec:xsec}. This means that after experimentally constraining the LD, the fitted values need to deviate less from the original values.

\subsection{Impact on inelastic gammas}
\label{sec:inel-gammas}

Experimental constraints on the HFB LD coming from double-differential cross-section data also improve the description of inelastic  gamma cross sections. Recent cross-section measurements of gamma transitions between different excited levels  provide complementary information to reaction cross sections. The accurate description  of inelastic gamma cross sections can be  challenging  from a theoretical perspective due to  structure issues and the many reaction  mechanisms involved.

\begin{figure*}[hptb]
 \centering
 \subfloat[Transition between states \#2 ($E_x$=846.8 keV) and \#1 (ground state).]{\label{fig:inel-gammas-2-1} \includegraphics[scale=0.68,keepaspectratio=true,clip=true,trim=0mm 0mm 0mm 0mm]{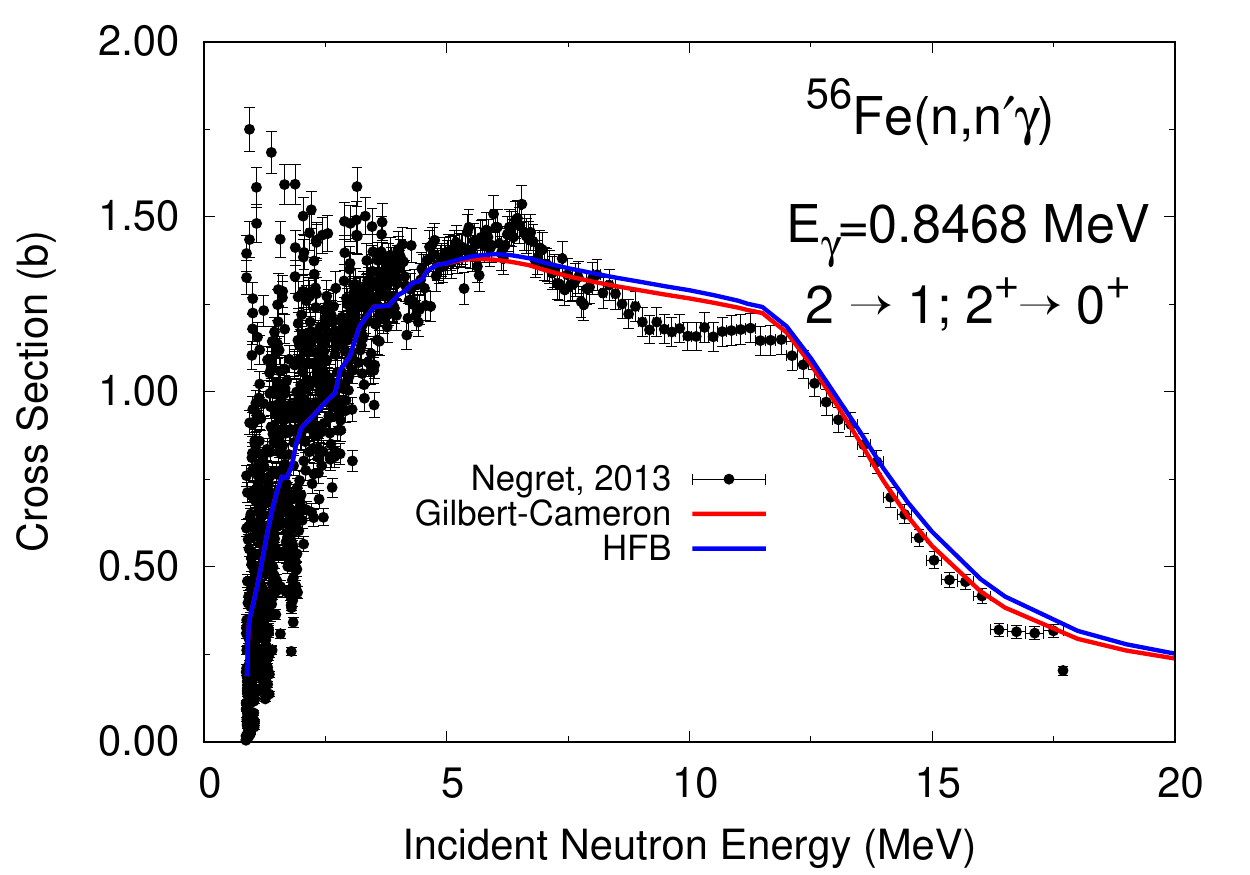}} \quad
 \subfloat[Transition between states \#5 ($E_x$ = 2.9415 MeV) and \#2 ($E_x$ = 846.8 keV).]{\label{fig:inel-gammas-5-2} \includegraphics[scale=0.68,keepaspectratio=true,clip=true,trim=0mm 0mm 0mm 0mm]{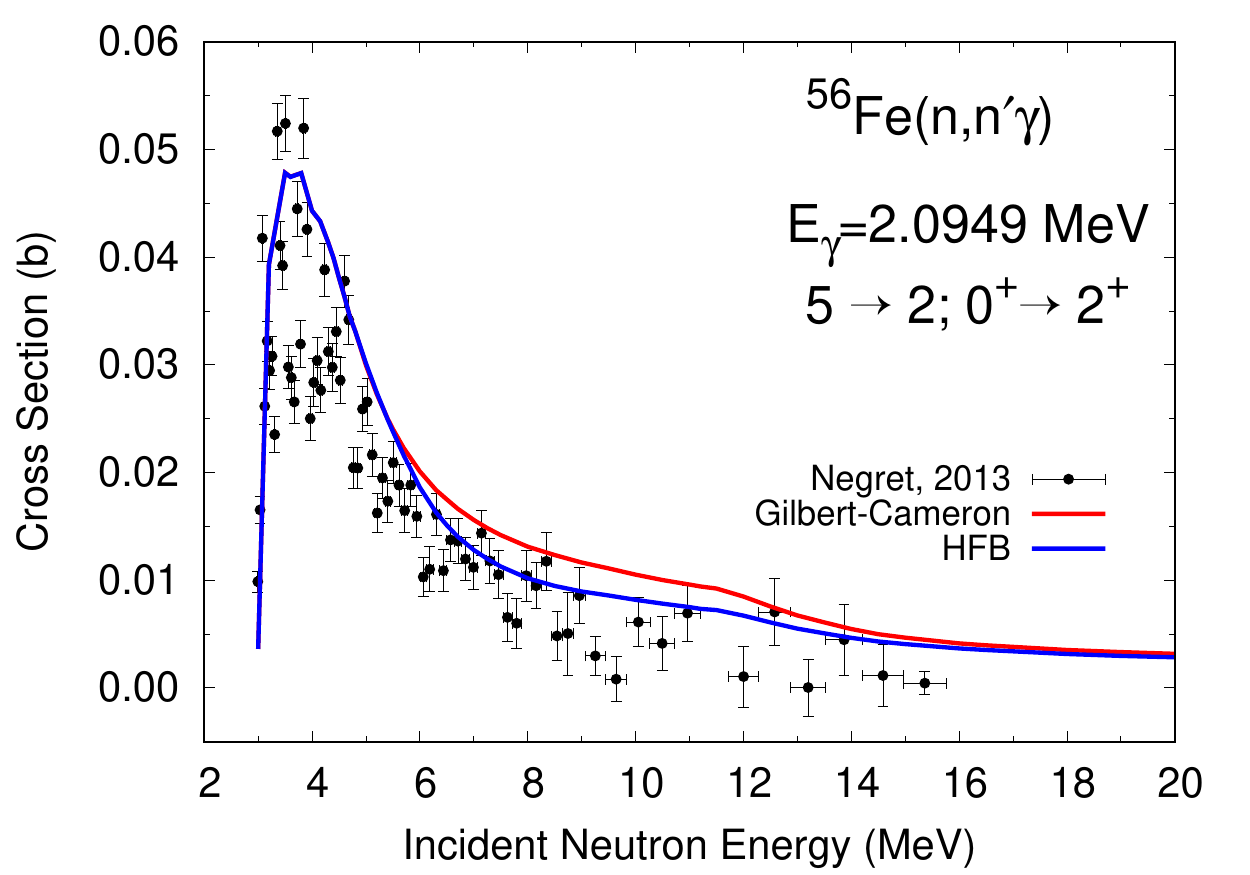}} 
 \\
 \subfloat[Transition between states \#7 ($E_x$ = 3.12011 MeV) and \#2 ($E_x$ = 846.8 keV).]{\label{fig:inel-gammas-7-2} \includegraphics[scale=0.68,keepaspectratio=true,clip=true,trim=0mm 0mm 0mm 0mm]{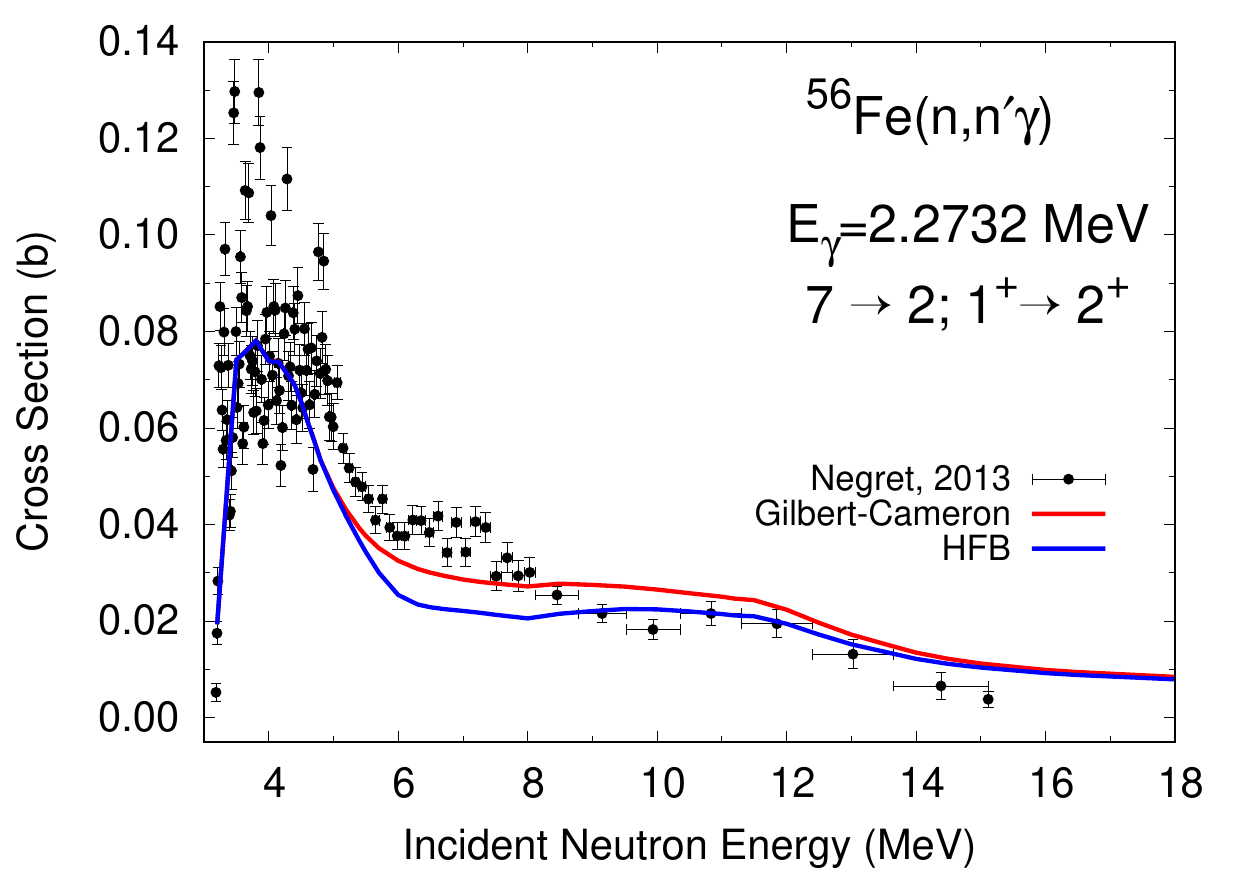}} \quad
 \subfloat[Transition between states \#31 ($E_x$ = 4.4477 MeV) and \#2 ($E_x$ = 846.8 keV); no data available.]{\label{fig:inel-gammas-31-2} \includegraphics[scale=0.68,keepaspectratio=true,clip=true,trim=0mm 0mm 0mm 0mm]{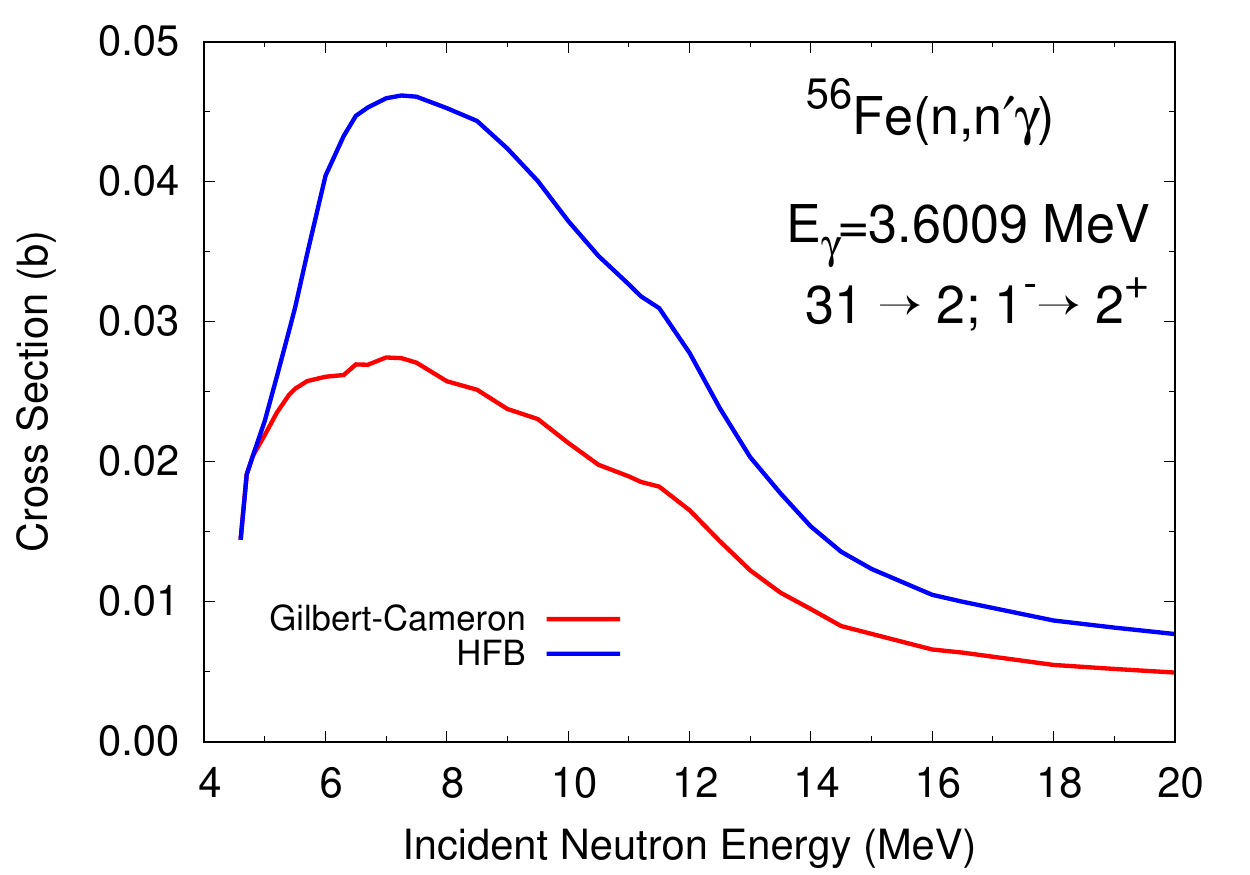}}
 \caption{Inelastic gamma cross sections for select transitions, as measured in Ref.~\cite{Negret-Iron}, with  model calculations using Gilbert-Cameron (red curves) and HFB (blue curve) LD models. }\label{fig:inel-gammas}
\end{figure*}

In Figure~\ref{fig:inel-gammas} we compare calculations of inelastic gamma cross sections obtained using the Gilbert-Cameron model (red curves) with the ones obtained using the HFB model with rescaled \nuc{56}{Fe} and  \nuc{56}{Mn} LD as described in Section~\ref{sec:spectra_and_LD} (blue curves). We have done this comparison for all transitions measured in the work of Negret et al. \cite{Negret-Iron}, and also other transitions that were not measured, but for brevity we selected only a few cases  in Figure~\ref{fig:inel-gammas}.  Figure~\ref{fig:inel-gammas-2-1} shows the gamma cross sections for the transition between states \#2 (first inelastic state with $E_x$=846.8keV) and \#1 (ground state). In this case the results are very similar. This is expected since this transition accounts for more than 95\% of the total inelastic   \cite{Negret-Iron}, thus most of the $\gamma$ transitions ultimately decay to this excited state before eventually reaching the ground state. Effects arising from the details of the LD models will  be more visible in transitions above the first inelastic state.

In Figure~\ref{fig:inel-gammas-5-2} we see the transition from level \#5 ($E_x$=2.9415MeV, $J^{\pi}=0^{+}$ state) to level \#2. In this example, as in many others not shown here, we can see a difference in the calculations and that the modified HFB model agrees better with experimental data. There are other transitions where differences are seen but it is difficult to determine which LD model is in better agreement. We show one such case in Figure~\ref{fig:inel-gammas-7-2} with  the gamma cross sections for the transition between level \#7 ($E_x$=3.12011MeV, $J^{\pi}=1^{+}$) to level \#2. Here,   the Gilbert-Cameron model for LD is closer to data between around 5 and 8 MeV, while above that the modified HFB is  in better agreement.

\begin{figure}[h]
 \centering
 \includegraphics[scale=0.70,keepaspectratio=true,clip=true,trim=0mm 0mm 5mm 0mm]{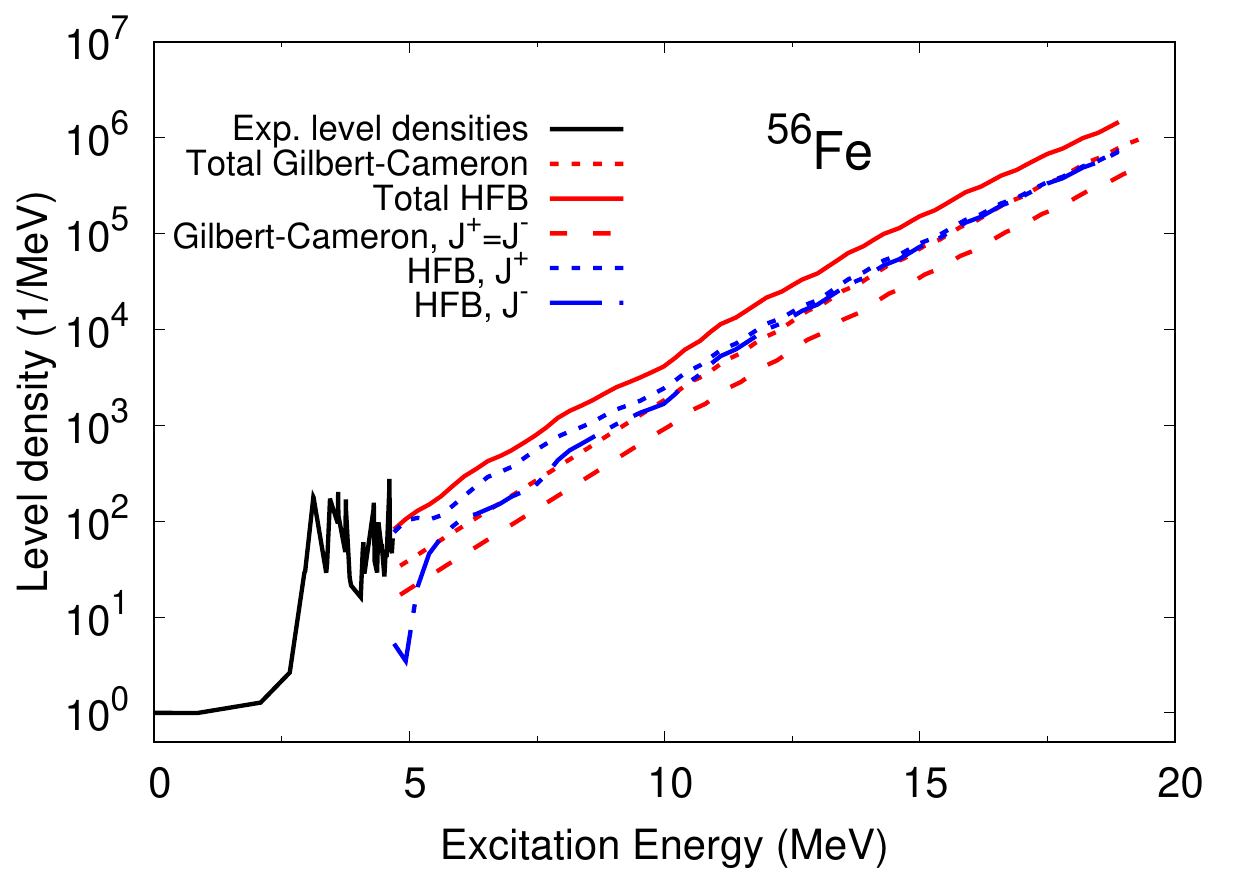}
 \caption{Assumed parity distribution from the Gilbert-Cameron and HFB LD models for \nuc{56}{Fe}.}\label{fig:parity-dist}
\end{figure}

Differences between calculations using HFB and Gilbert-Cameron, although generally favoring the microscopic approach, are not too big.  
However, in cases like the one in Figure~\ref{fig:inel-gammas-31-2}, which shows the transition between levels \#31 ($E_x$=4.4477MeV, $J^{\pi}=1^{-}$) and \#2, we see a large difference between the predicted gamma cross sections from the two different models. Noting that here we have a transition between a negative-parity state to a positive one, this large difference can likely be attributed to the fact that the HFB model has independent level and spin distributions for each parity value, while the phenomenological Gilbert-Cameron assumes   equal parity distributions (see Figure~\ref{fig:parity-dist}).  As we can see, there are no measurements for this transition. However, due to the fact that the HFB is more fundamental in its microscopic nature, with more realistic spin and parity distributions, and has been modified keeping internal consistency, its predictions should be more credible than those of the Gilbert-Cameron model.
New experimental results for the gamma-decay of negative parity states in \nuc{56}{Fe} would be very helpful to confirm parity distribution in  \nuc{56}{Fe}. Considering that the incomplete picture of known levels hinders the reliable determination of spin and parity distributions, as shown in Section~\ref{sec:spin-dist}, different theoretical approaches such as those of Refs.~\cite{Horoi:2003,Alhassid:2007,Mocelj:2007} can provide valuable information to improve the prediction of inelastic gamma cross sections. This will be investigated in a future work.  There could also be an impact due to a better modeling of the direct and pre-equilibrium process as the one mentioned in Ref.~\cite[Figures~7 and~8]{Dupuis:2015}, or to the use of different gamma strengths \cite{Goriely:2019}.

\section{Sensitivity studies}
\label{sec:sensitivity}

Even though in the particular case studied here the experimental data from double-differential spectra, as well as ($n,p$) data, point towards smaller oscillations  in the LD, it does not necessarily rule them out. Some structures are seen in spectra data and the LD in an extended region of excitation energies may affect the cross section in the same incident energy region for a particular reaction. Therefore, a change of position and shape of the structures in LD can have similar impact in the cross sections as the rescaling shown in Section~\ref{sec:spectra_and_LD}. In order to quantify  this effect, we performed sensitivity studies correlating changes in \nuc{56}{Fe}($n,p$) cross section  to changes in \nuc{56}{Fe} and \nuc{56}{Mn} HFB LD at specific excitation energies. 
For this we define the fractional variation $F$ for a specific channel as:
\begin{equation}
\label{eq:sensitivity}
F(E_{\mathrm{inc}},E^{\prime}_{x}) = \frac{\sigma_{\mathrm{up}}(E_{\mathrm{inc}}) - \sigma_{\mathrm{down}}(E_{\mathrm{inc}})}{2\sigma_{0}(E_{\mathrm{inc}})},
\end{equation} 
where 
$\sigma_{\mathrm{up/down}} $ are the cross sections calculated with a modified total LD (i.e.,\ the sum of positive and negative parities) $\rho^\prime_{\mathrm{up/down}}(E_{x},E^{\prime}_{x})$. This modified LD is rescaled up or down by a constant factor $\Delta\rho$ only at  $E^{\prime}_{x}$ and remains unmodified everywhere else. The mathematical details of how this is done, especially considering the finite excitation-energy grid in which LD are used in numerical calculations, can be found in Appendix~\ref{app:Smat}.
The central cross section $\sigma_{0}(E_{\mathrm{inc}})$ do not have any up/down variation in any LD. 
In the results to follow we adopt a LD variation of $\Delta\rho = 30\%$. As we detail in Appendix~\ref{app:Smat}, the fractional variation is directly related to sensitivity matrices and covariances, allowing one relate covariances in LD to those in the cross-section experimental data.

\begin{figure}[hptb]
\centering 
\subfloat[Fractional variation of \nuc{56}{Fe}($n,p$)  relative to changes in \nuc{56}{Fe} LD.]{\label{fig:sens-fe56-MT103} \includegraphics[scale=0.80,keepaspectratio=true,clip=true,trim=10mm 0mm 10mm 0mm]{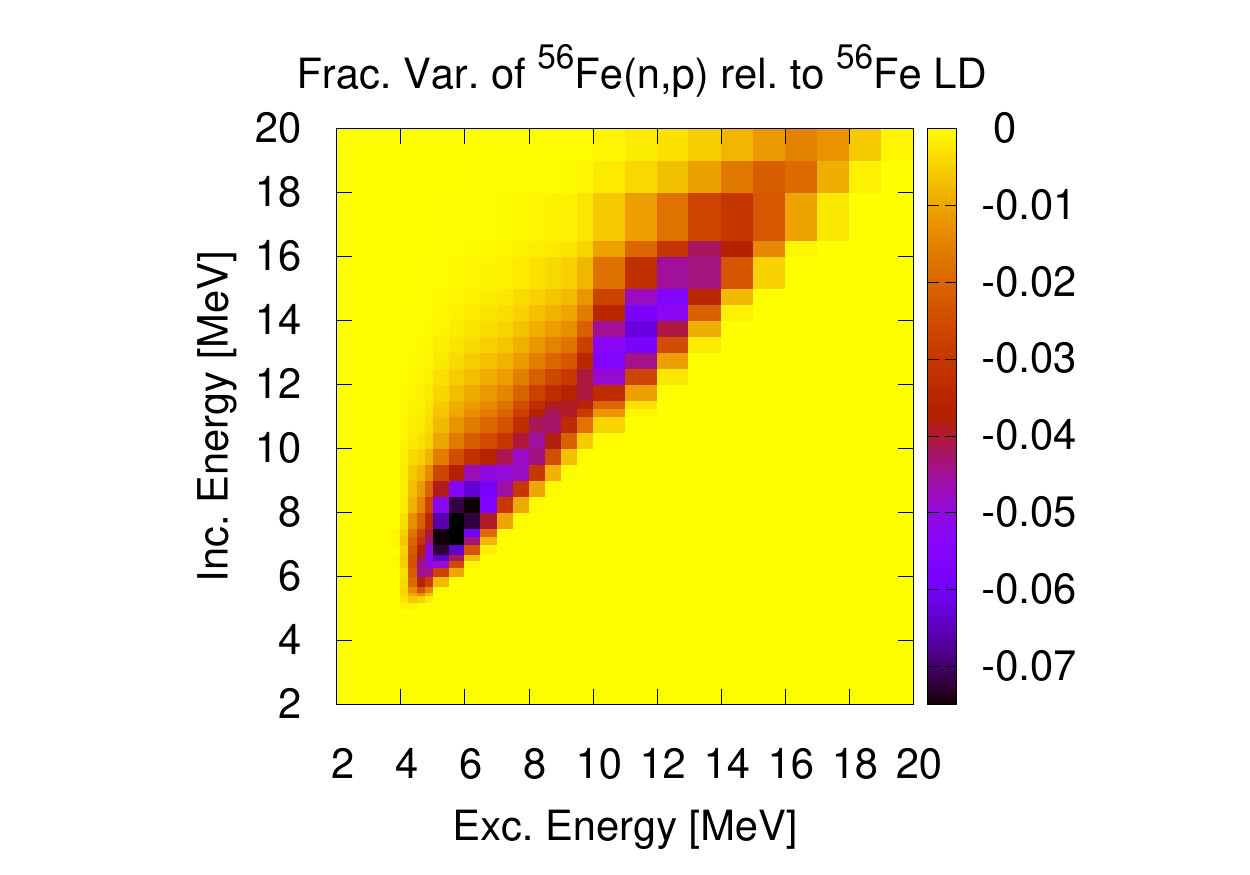}}
\\
\subfloat[Fractional variation of \nuc{56}{Fe}($n,p$) relative to changes in  \nuc{56}{Mn} LD variations.]{\label{fig:sens-mn56-MT103} \includegraphics[scale=0.80,keepaspectratio=true,clip=true,trim=10mm 0mm 10mm 0mm]{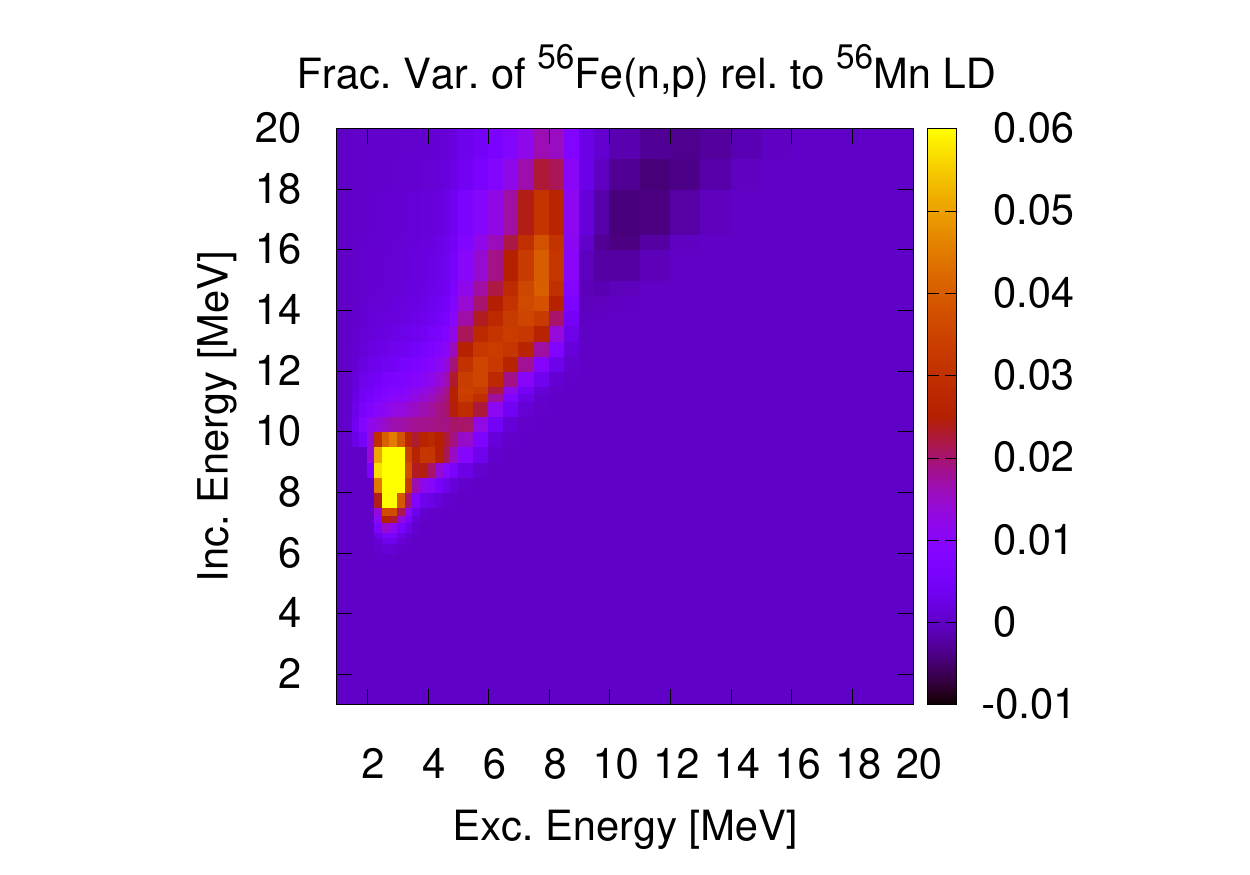}}

\caption{Fractional variations of  \nuc{56}{Fe}($n,p$) cross sections, at a given incident energy, relative to changes in the \nuc{56}{Fe} (upper panel) and \nuc{56}{Mn} (bottom panel) LD  at specific excitation energies. }\label{fig:sensitivities-MT103}
\end{figure}

In Figure~\ref{fig:sensitivities-MT103} we show the fractional variations of \nuc{56}{Fe}($n,p$) relative to changes in the LD for the target (\nuc{56}{Fe}, Figure~\ref{fig:sens-fe56-MT103}) and ($n,p$) residual (\nuc{56}{Mn}, Figure~\ref{fig:sens-mn56-MT103}) nuclei,  as functions of both the neutron incident energy and the excitation energy at which the LD is given. For completeness we also analyzed the sensitivities associated with LD variations in the compound nucleus \nuc{57}{Fe} and, as it would be expected, the ($n,p$) cross sections are much less sensitive to \nuc{57}{Fe} LD, when compared to \nuc{56}{Fe} and \nuc{56}{Mn}, hence we do not show the corresponding plot.

Looking at Figures~\ref{fig:sens-fe56-MT103} and~\ref{fig:sens-mn56-MT103} we see that the fractional variations are spread-out in the ($E_{\mathrm{inc}},E_{x}$) plane around peaks and valleys of sensitivity. This means that the ($n,p$) cross section at a given incident energy is affected by LD at a certain extended region of excitation energy. Moreover, the regions in the cross sections that are most sensitive to variations in the LD are around $E_{x}$ = 6 MeV and 12 MeV for the \nuc{56}{Fe} LD (Figure~\ref{fig:sens-fe56-MT103}), and at $E_{x}$ = 3 MeV and a wider peak between around 6 and 9 MeV for the \nuc{56}{Mn} LD (Figure~\ref{fig:sens-mn56-MT103}).  The sharp cutoffs seen at low excitation energies in Figure~\ref{fig:sensitivities-MT103} are due to the fact that below those energies discrete levels are used in the calculations instead of LD.

The connection between LD and cross-section allows us to verify the existence and intensity of LD structures predicted by fundamental models like the HFB by examining experimental reaction data. 
This is often overlooked in applications as phenomenological LD models assume energy-dependent smooth functionals, even at lower excitation energies when some structure coming from discrete levels should be expected. 
Additionally,  even when indirect measurements of LD are made (e.g. using the Oslo method~\cite{Schiller2000}) showing the existence of structure in the LD, these data are fitted to smooth model functionals before being applied to reaction calculations (e.g., Ref.~\cite{Spyrou2014}). The sensitivity results presented here are qualitatively consistent with the sensitivities presented in Ref.~\cite{Rauscher:2012}, even though Ref.~\cite{Rauscher:2012} varies LD at the separation energy while we focus on an extended region of LD.
This analysis of sensitivities and correlations can be extended to 
energy spectra. However, this analysis becomes more complicated by the added dimensionality. 

In addition to provide important scientific insights into the details of the LD constrained by differential and integral cross-section data, the LD sensitivities can serve as direct input for fitting within any Bayesian approach (e.g. KALMAN code \cite{KALMAN}). This may allow reaction evaluators to describe even the minor details and structures observed in the neutron  spectra and  cross sections such as ($n,p$), ($n$,$\alpha$), ($n,2n$), etc. 
%
Additionally, one can reverse the flow of probability to use measured experimental reaction data to inform the LD along the way outlined in Appendix~\ref{app:Smat}.
In this work, we have presented sensitivities by varying the total LD, which means that we have kept the positive-to-negative parity ratio constant. However, we did perform exploratory studies on parity-dependent sensitivities and we were able to separate the impacts in cross sections coming from the model-assumptions for the different parities. Again, this can provide significant assistance in the development of microscopic models for LD.


As we mentioned in Section~\ref{sec:LD_models}, the starting point of this work was the development of ENDF/B-VIII.0 evaluation for \nuc{56}{Fe}. As that work was concluding, it became known that the main experimental set that underpinned the total inelastic reaction cross section, namely Nelson et al. \cite{Nelson2005}, should have been normalized 11.8\% lower.   
At some point in the future, a new evaluated file should be released to rectify this.  
However, we do not expect this to change any conclusion or qualitative result of the present work.
The major impact in the evaluated inelastic cross sections should be in the plateau region (see Figure~9 of Ref.~\cite{CIELO-IRON}).   This is where neutron incident energies range between $\approx$5 and $\approx$11 MeV and where the relative importance of the inelastic channel is the greatest, 
below or just around the ($n,p$) threshold. 

\begin{figure}[hptb]
\centering
\subfloat[Fractional variation of \nuc{56}{Fe}($n$,inel) relative to changes in  \nuc{56}{Fe} LD.]{\label{fig:sens-fe56-MT4} \includegraphics[scale=0.80,keepaspectratio=true,clip=true,trim=10mm 0mm 10mm 0mm]{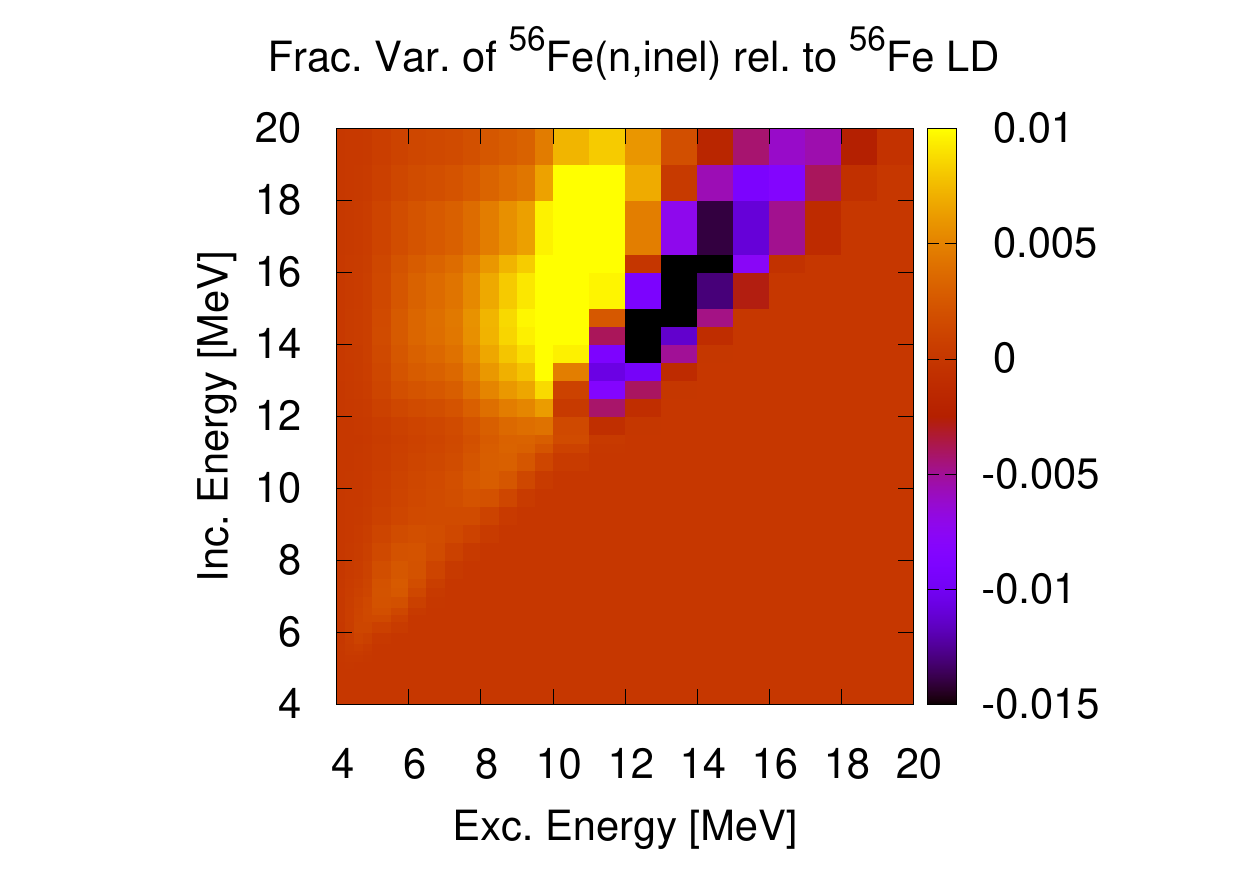}}
\\
\subfloat[Fractional variation of \nuc{56}{Fe}($n$,inel) relative to changes in \nuc{56}{Mn} LD.]{\label{fig:sens-mn56-MT4} \includegraphics[scale=0.80,keepaspectratio=true,clip=true,trim=10mm 0mm 10mm 0mm]{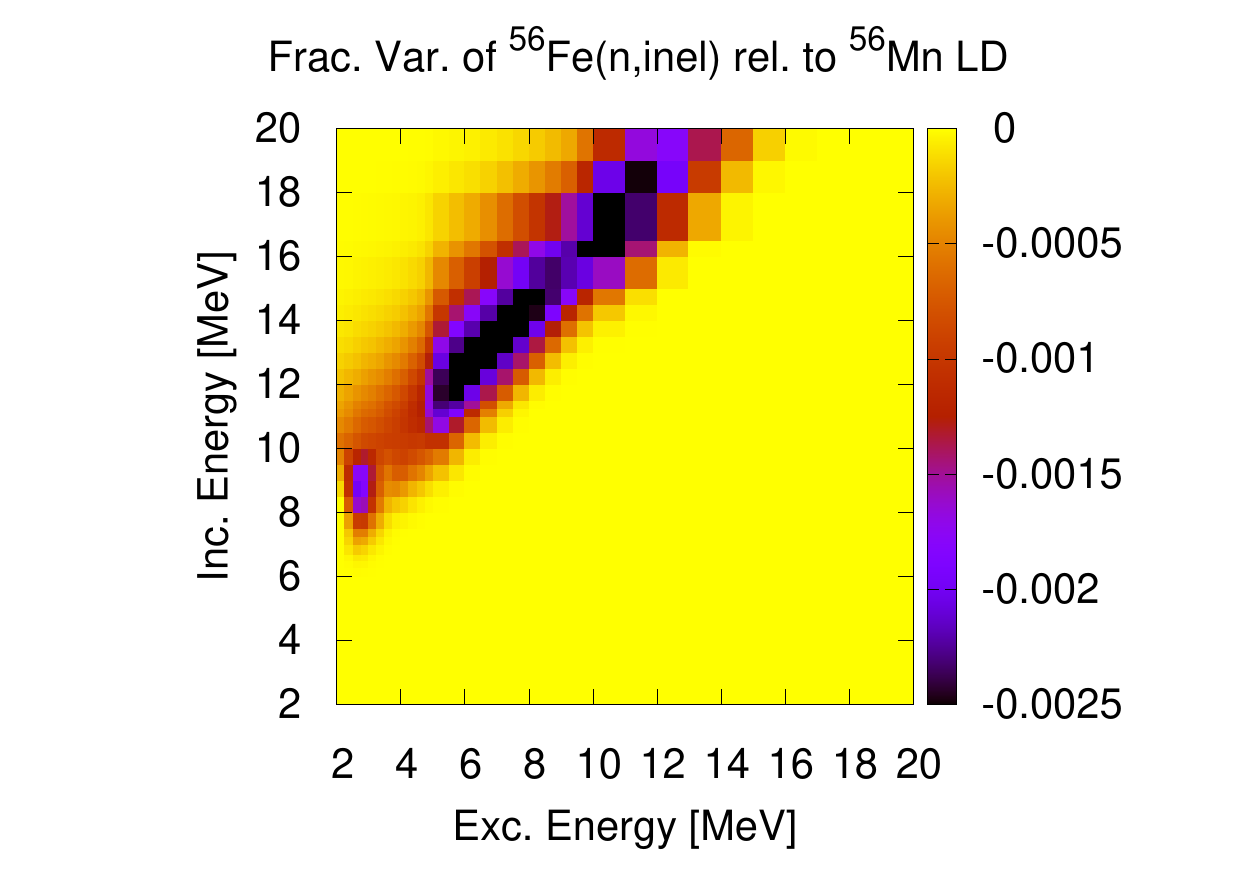}}
\caption{Fractional variations of \nuc{56}{Fe}($n$,inel) cross sections, at a given incident energy, relative to changes in the \nuc{56}{Fe} (upper panel) and \nuc{56}{Mn} (bottom panel) LD at specific excitation energies.}\label{fig:sensitivities-MT4}
\end{figure}

To confirm this, we calculated the fractional variations of the  inelastic channel relative to the \nuc{56,57}{Fe} and \nuc{56}{Mn} LD, as can be seen in Figure~\ref{fig:sensitivities-MT4} (again, sensitivities for \nuc{57}{Fe} LD are too small to be shown).  They clearly show that such sensitivities are overall much smaller than for the \nuc{56}{Fe}($n,p$)\nuc{56}{Mn} reaction. 
The only case where the order of magnitude of inelastic-channel fractional variation is comparable to the ($n,p$) ones is for \nuc{56}{Fe} LD (Figure~\ref{fig:sens-fe56-MT4}), but even so, they are quite small in the region where inelastic cross sections will change, becoming larger only at higher incident energies, perhaps due mostly to the  competition with the  ($n,2n$) channel. 

The differences in sensitivities for different reactions at a given energy range reflect different aspects of the reaction channels. For instance, at incident energies below 10 MeV the only channels in competition with inelastic are neutron capture, which is small, and elastic. Therefore, the inelastic channel exhausts most of the absorption cross section and has no possibility of growth. Around 8-9 MeV, however,  the ($n,p$) cross section becomes large  enough to make some room for changing inelastic. This happens in spite of the fact that,  even in its peak, ($n,p$) is about ten times smaller than the inelastic plateau. Between excitation energies of around  10 and 12 MeV, inelastic sensitivities (Figure~\ref{fig:sensitivities-MT4}) are positive and then abruptly change to negative. This is possibly related to gamma-emission channels and/or to the fact that increasing LD at $E_x\approx$ 10-12 MeV increases the population of \nuc{56}{Fe} continuum. At these energies gammas leading to ($n,n\gamma$) can still compete with emission of the second neutron (n,2n). If the LD at higher energies increases, the  population of \nuc{56}{Fe} continuum is shifted to higher energies which favors second neutron emission. This could serve as guidance to find relatively minor changes in target-nucleus LD   around the ($n,2n$) threshold which would allow to adjust the  ($n,n$)/($n,2n$) ratio in reaction evaluations.

\section{Summary and conclusions}
\label{Sec:Conclusion}

We have discussed that phenomenological level-density (LD) models assume simplifications and approximations which are only loosely constrained by experimental data. 
The constraints provided by $D_0$ and/or $D_1$, when they are available, is insufficient as it only fixes the LD at  the neutron separation energy of the compound nucleus.  
This leaves the rest to be described by functionals which at best are insensitive to structure in the LDs and at worst are  
stretched beyond reasonability in order to optimize the cross-section agreement with experimental data.

We have demonstrated that by starting off with a microscopic, more predictive LD model, one can use experimental data from neutron spectra to constrain and rescale the structures in HFB LD in an extended region of excitation energy. This leads to a more self-consistent framework in which LDs that agree with observed cumulative-level distribution also agree with measured cross sections. Additionally, the more realistic parity and spin distributions provide better agreement with measured inelastic gamma cross section and increase reliability of the predicted ($n,n^\prime\gamma$) when no data is available, especially in cases involving 
unbalanced parity distributions.

We have also analyzed sensitivity matrices connecting variations in LD at a given excitation energy to cross-section changes at a given incident energy. 
This allowed us to observe peaks and valleys of sensitivity, indicating that some excitation-energy regions of the LD impact cross sections  more than others. Turning this around, cross-section data can constrain specific regions of LD, leading to more realistic and predictive LD models and reaction calculations. This may lead to structures in the LD, or at least test predicted structures, and thus estimate how realistic are the assumptions made in fundamental models like HFB.
Furthermore, these LD sensitivity matrices can serve as inputs for cross-section fitting, in principle allowing to describe detailed structures observed in  spectra and  cross-section data, being a powerful additional tool for reaction evaluators.

Special attention was devoted to the \nuc{56}{Fe}($n,p$) reaction, which was used as test case of our approach. It was shown that by using cross-section data to constrain a more fundamental set of LD we  improve the agreement with the precise ($n,p$) data while consistently producing \nuc{56}{Mn} LD that are in agreement with observed discrete levels, as well as more reliable inelastic gamma cross sections. It was also shown how LD/cross-section sensitivities can be used as an evaluation tool to describe details of precise reaction measurements.

The results presented here serve as an important guidance coming directly from experimental cross-section measurements, constraining LD not only at separation energy but rather at an extended range of excitation energy.

\appendix
\section{Sensitivity matrices}
\label{app:Smat}

Both the ENDF-6~\cite{ENDF-6} format and the EMPIRE reaction code describe the computed cross sections,  model parameters and level densities using linear interpolation.  Interpolated functions such as the cross section at a given incident energy may be written using a spline basis:
\begin{equation}
	\sigma(E) = \sum_m \sigma_m B_m(E).
	\label{eq:sigmaspline}
\end{equation}
Here we define $\vec{\sigma}=\{\sigma_1, ..., \sigma_{M}\}$, where $\sigma_m=\sigma(E_m)$ and $B_n(E)$ are ``triangular'' functions so that Eq. \eqref{eq:sigmaspline} is a linear spline representation of the cross section \cite{Wahba}.  Similarly,
\begin{equation}
	\rho(E_x) = \sum_n \rho_n B_n(E_x)
		\label{eq:rhospline}
\end{equation}
define $\vec{\rho}=\{\rho_1, ..., \rho_{N}\}$ where $\rho_n=\rho(E_{x,n})$, and thus $\vec{\sigma}(\vec{\rho})$.  

We define the sensitivity matrix as
\begin{equation}
	S_{ij}=\partial \sigma_i/\partial \rho_j
\end{equation}
which has units of area times energy, e.g. barns $\times$ MeV if $\sigma$ has units of barns and $\rho$ has units of 1/MeV.  Here we consider for simplicity only the total level density, but the spin/parity dependency, or any other parameter dependency, of the  level density could also be made explicit.

\subsection{Variations}
Consider a small variation in the $i^{\mathrm{th}}$ element of $\vec{\rho}$, $\delta\rho_i$.  This corresponds to a variation in the level density of $\Delta\rho(E_x)=\delta\rho_iB_i(E_x)$ in our linear spline basis.  Note variations of this form can easily be recast as an energy dependent normalization factor.   In terms of the sensitivity matrix, this variation leads to a variation of cross section coefficients of  $\delta\sigma_i=S_{ij}\delta\rho_j$.  This is equivalent to a spline basis variation of 
\begin{equation}
	\Delta\sigma(E)=\sum_{ij}S_{ji}\delta\rho_iB_j(E)
\end{equation}
In~Eq. \ref{eq:sensitivity}, the fractional variation is then
\begin{equation}
	F(E)=\frac{\sigma_{\mathrm{up}}(E) - \sigma_{\mathrm{down}}(E)}{2\sigma_{0}(E)} = \frac{\Delta\sigma(E)}{\sigma(E)}
\end{equation}
If we evaluate this at the spline points $E_i$, we see that the fractional variation is directly related to the sensitivity matrix:
\begin{equation}
	F_i=\frac{\Delta\sigma(E_i)}{\sigma(E_i)} = \sum_i\frac{S_{ji}\delta\rho_i}{\sigma_j}.
\end{equation}

\subsection{Covariance propagation}
The final probability distribution for the cross section $P(\vec{\sigma})$ depends on the probability distribution assumed for the level density parameters through
\begin{equation}
P(\vec{\sigma})=\int d\vec{\rho} P(\vec{\sigma}|\vec{\rho}) P(\vec{\rho}),
\label{eq:forwardUQ}
\end{equation} 
where $P(\vec{\sigma}|\vec{\rho})$ is the conditional probability of $\vec{\sigma}$ given $\vec{\rho}$.  
With this we can forward propagate uncertainty from the level density to the cross section.

In practice, this conditional probability is a delta function, $P(\vec{\sigma}|\vec{\rho})=\delta(\vec{\sigma}-\vec{\sigma}(\vec{\rho}))$.  If we assume that the probability distributions for the cross section $P(\vec{\sigma})$ and level density $P(\vec{\rho})$ are multivariate normal distributions and completely therefore characterized by the mean values and corresponding covariances, then we have a Gaussian Process Regression model \cite{Rasmussen} of the cross section.  
Assuming the variations from the mean values are small, we can use Eq.\ \eqref{eq:forwardUQ} to determine the final covariance of the cross section using the so-called ``sandwich formula'':
\begin{equation}
	\Delta^2\sigma_{ij}=\sum_{kl} S_{ik}\Delta^2\rho_{kl} S_{jl}.
\end{equation}
Using the linear spline basis, we can compute the Kriging estimate \cite{Rasmussen} of the cross section covariance between energies $E$ and $E'$ as 
\begin{eqnarray}
	\Delta^2\sigma(E,E') & = & \sum_{ij} B_i(E) \Delta^2\sigma_{ij} B_j(E')\\
	& = & \sum_{ijkl} B_i(E) S_{ik}\Delta^2\rho_{kl} S_{jl} B_j(E')
	\label{eq:kriging}
\end{eqnarray}

\subsection{Likelihood back-propagation}
Using Bayes' theorem \cite{Pearl}, 
\begin{equation}
	L(\vec{\rho}|\vec{\sigma}) = P(\vec{\rho}|\vec{\sigma}) = \frac{P(\vec{\sigma}|\vec{\rho})P(\vec{\rho})}{P(\vec{\sigma})}
\end{equation}
we may ``reverse the flow'' of probability and use measured cross section data to constrain the level densities. 
Here the likelihood  $L(\vec{\rho}|\vec{\sigma})$ is just the probability of $\vec{\rho}$ given $\vec{\sigma}$.  

Again, assuming that all probability distributions are characterized by the mean value of the cross section and its corresponding covariance, we have 
\begin{equation}
	\Delta^2\rho_{ij}=\sum_{kl} \tilde{S}_{ik}\Delta^2\sigma_{kl} \tilde{S}_{jl}.
\end{equation}
These modified sensitivity matrices are $\tilde{S}_{ij} = \partial \rho_i/\partial \sigma_j = (\partial \sigma_j/\partial \rho_i)^{-1}$.  As in Eq. \eqref{eq:kriging}, we can construct 
\begin{equation}
	\Delta^2\rho(E_x,E_x') 
	 =  \sum_{ijkl} B_i(E_x) \tilde{S}_{ik}\Delta^2\sigma_{kl} \tilde{S}_{jl} B_j(E_x')
	\label{eq:kriging2}
\end{equation}
Thus, we have used the likelihood to back-propagate the covariance and to inform the level density. In this way we can quantify level-density uncertainties, in the whole excitation-energy range in a way directly constrained the uncertainties in cross-section measurements.

\section*{Acknowledgments}

The work at Brookhaven National Laboratory was sponsored by the Office of Nuclear
Physics, Office of Science of the U.S. Department of
Energy under Contract No. DE-AC02-98CH10886 with
Brookhaven Science Associates, LLC.
This work was performed under the auspices of the National Nuclear Security Administration of the U.S. Department of Energy at Los Alamos National Laboratory under Contract No. 89233218CNA000001.

\bibliographystyle{apsrev4-1} 
\bibliography{leveldens-HFB}

\end{document}